\documentclass[prb,  twocolumn, showpacs, superscriptaddress,longbibliography]{revtex4-1}
\usepackage{amsmath, amsfonts, amssymb, mathrsfs}
\usepackage{graphicx, color}

\usepackage[caption=false]{subfig}
\usepackage{bm}
\usepackage{braket}
\usepackage{epstopdf}
\usepackage{dsfont}
\usepackage[dvipsnames]{xcolor}
\usepackage{program}
\usepackage{hyperref}
\hypersetup{
   colorlinks,
   citecolor=blue,
   filecolor=blue,
   linkcolor=blue,
   urlcolor=blue,
   breaklinks=true
}

\graphicspath{Graphes}

\newcommand{\Kappa}{\mathcal{K}}

\definecolor{orange}{rgb}{1,0.5,0}

\definecolor{grey}{rgb}{.6,.6,.6}

\newcommand{\ti}[1]{{\textit{#1.}}}

\begin{document}

\title{Defining a bulk-edge correspondence for non-Hermitian Hamiltonians via singular-value decomposition}

\author{Lo\"{i}c Herviou}
\affiliation{Department of Physics, KTH Royal Institute of Technology, Stockholm, 106 91 Sweden}
\author{Jens H.~Bardarson}
\affiliation{Department of Physics, KTH Royal Institute of Technology, Stockholm, 106 91 Sweden}
\author{Nicolas Regnault}
\affiliation{Laboratoire de Physique de l'Ecole Normale Sup\'erieure, ENS, Universit\'e PSL, CNRS, Sorbonne Universit\'e, Universit\'e Paris-Diderot, Sorbonne Paris Cit\'e, Paris, France}

\begin{abstract}
We address the breakdown of the bulk-boundary correspondence observed in non-Hermitian systems, where open and periodic systems can have distinct phase diagrams. 
The correspondence can be completely restored by considering the Hamiltonian's singular-value decomposition instead of its eigendecomposition. 
This leads to a natural topological description in terms of a \emph{flattened singular decomposition}. 
This description is equivalent to the usual approach for Hermitian systems and coincides with a recent proposal for the classification of non-Hermitian systems.
We generalize the notion of the entanglement spectrum to non-Hermitian systems, and show that the edge physics is indeed completely captured by the periodic bulk Hamiltonian. 
We exemplify our approach by considering the chiral non-Hermitian Su-Schrieffer-Heger and Chern insulator models.
Our work advocates a different perspective on topological non-Hermitian Hamiltonians, paving the way to a better understanding of their entanglement structure.
\end{abstract}
\date{\today}
\maketitle

\section{Introduction}
In recent decades, topology has become a fundamental concept in condensed matter physics\citep{Kane2005, Fu2007, Fu2007-2, Hasan2010, ShenBook, BernevigBook}.
A comprehensive classification of different topological phases under various sets of symmetries led to vast advances in our understanding of electronic systems, in particular for closed systems described by Hermitian Hamiltonians\citep{Schnyder2008, Kitaev2009, Chiu2016, Kruthoff2017, Bradlyn2017, Cano2018}.
Indeed, topology explains the existence and resilience of numerous physical properties (such as unconventional edge or surface states) and provides a unified description of unconventional phases and phase transitions. 
One of the key ideas and guiding principles in topological matter is the bulk-boundary correspondence\citep{BernevigBook, AsbothBook, OrtmannBook}: nontrivial topological invariants in the bulk of a system directly translate into gapless edge physics.
This correspondence has been verified in a plethora of models, including even higher order topological phases\citep{Benalcazar2017, Schindler2018}.

It is therefore not surprising that attempts have been made to extend those concepts to non-Hermitian models\citep{Gong2018, Liu2018}.
Non-Hermitian Hamiltonians provide a simple, albeit restricted, description of open systems.
Instead of considering the full and overly complex problem of microscopically modeling a system coupled to its environment, or of working in a Lindblad formalism which governs the time evolution of density matrices, we can model dissipative memory-less environments by breaking the Hermiticity of the Hamiltonian\citep{ChuangBook}.
This simplified approach has successfully described numerous experiments and phenomena, with applications ranging from mechanical and optical metamaterials\citep{Lu2014, Parto2018, Takata2018, Zhu2018, Longhi2018} to heavy-fermions systems\citep{Kozii2017, Yoshida2018}.
The topological properties observed in these systems can significantly differ from their Hermitian counterparts\citep{MartinezAlvarez2018}, and several questions about their fundamental applicability remain open---the validity of the bulk-boundary correspondence being one of them\citep{Lee2016, Leykam2017, Xiong2018, MartinezAlvarez2018, Yao2018, Yao2018-2, Kunst2018, Lee2018-3, Jin2018, Edvardsson2018}.
Indeed, it has been shown that in several models the phase diagram can strongly depend on the boundary conditions, where even the bulk spectra change depending on whether one considers periodic or open systems in sharp contrast with topological Hermitian systems.
Similarly, the existence and stability of edge states in such models have been questioned\citep{Xiong2018, Kunst2018, Lee2018-3}.

In this work, we provide a simple mathematical explanation for why the bulk-boundary correspondence breaks down in non-Hermitian systems, and propose a change of paradigm in the way we look at and define topology in these systems.
For a topological classification of non-Hermitian matrices to make physical sense, with resilience to small perturbations, and for any bulk-boundary correspondence to stand, it is more fruitful to think in terms of singular values of the Hamiltonian than in terms of eigenvalues.
Moreover, in contrast to the eigenvalues, the singular values are well-behaved in the thermodynamic limit.
Below, through the example of the chiral non-Herminitian Su-Schrieffer-Heger (nH-SSH) model\citep{SSH1980, Yao2018-2, Yin2018}, we illustrate concretely where the usual eigenvalue-based topology fails for non-Hermitian systems, and how standard results are phenomenologically recovered from the singular value decomposition (SVD).
We then formalize a topological description for non-Hermitian models, and give explicit examples of topological invariants in different symmetry classes and different dimensions.
Finally, we introduce an analog to the entanglement spectrum for the nH-SSH model, and show that the bulk-boundary correspondence is indeed recovered.
We verify that this analog also applies to higher-dimensional models.
\\

\ti{Notation}---We consider a matrix $H$. We denote with $E$ and $P$ its eigenvalue decomposition, while its singular value decomposition is the set of three matrices $U$, $V$ and $\Lambda$ such that:
\begin{equation}
H=P E P^{-1}=U \Lambda V^\dagger.
\end{equation}
$\Lambda$ is a diagonal and real positive matrix, whose eigenvalues are the so-called singular values, and $U$ and $V$ are two unitary matrices.
The columns of $U$ and $V$ are the left and right singular vectors. 
While the decomposition admits some gauge freedom, both $\Lambda$ (up to ordering) and $Q = U V^\dagger$ are uniquely defined.
The two decompositions are similar in a Hermitian setting: $\Lambda=\lvert E \rvert$ and one can choose $U=P$, $V^\dagger=\text{sgn}(E) P^\dagger$. 
This property breaks down for non-Hermitian matrices, with one crucial exception: each zero singular value corresponds to one Jordan block with zero eigenvalue.

\section{Breakdown of the bulk boundary correspondence}

\subsection{Bulk-boundary correspondence}
\begin{figure}
\includegraphics[width=\columnwidth]{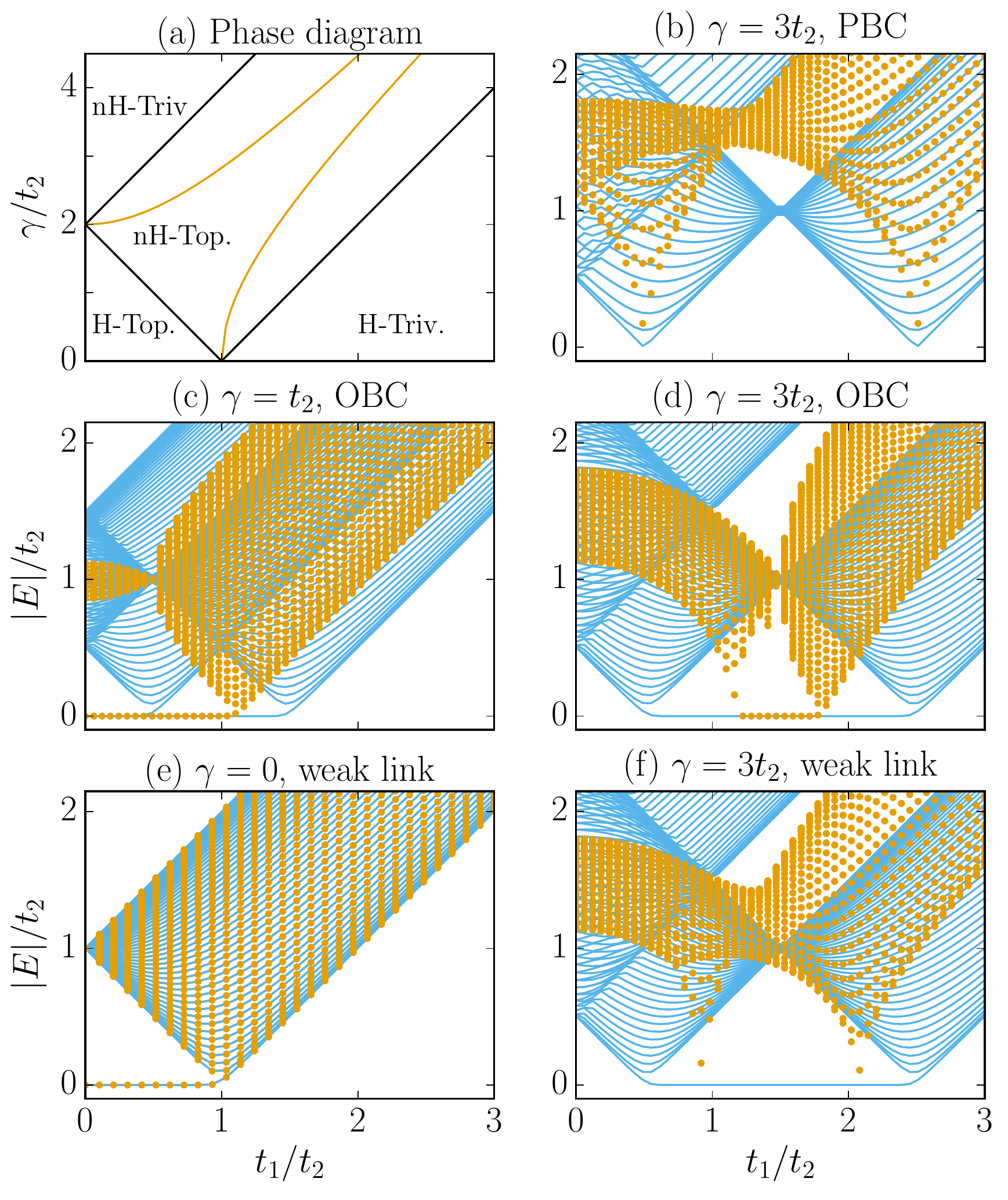}
\caption{(a) Superposition of the phase diagrams of the non-Hermitian SSH model. Black lines mark gap closings for energies with PBC, and for singular values both with PBC and OBC. Orange lines mark gap closings for energies with OBC. In the PBC case, we observe four different phases: the Hermitian topological and trivial phase are simple extension of the phases of the Hermitian SSH model. (b-f) Singular values (blue lines) and absolute value of energies (orange dots) for different $\gamma$ and boundary conditions. (b-d) The mismatch between singular values and energies for OBC, and the breakdown of bulk-boundary correspondence for energies are apparent. (e) Hermitian and (f) non-Hermitian SSH model with OBC and a weak link $10^{-12}t_2$ connecting the two edges, for $L=50$. The edges states acquire a macroscopic energy in the non-Hermitian topological phase, while the Hermitian phase is essentially unaffected.
}
\label{fig:PD}
\end{figure}

The bulk-boundary correspondence breaks down in certain non-Hermitian models\citep{Lee2016, Leykam2017, Xiong2018, MartinezAlvarez2018, Yao2018, Yao2018-2, Kunst2018}, an effect dubbed the non-Hermitian skin-effect. 
As an illustration, let us define the chiral nH-SSH model\citep{SSH1980, Yao2018-2, Yin2018} as:
\begin{align}
H_\textrm{nH-SSH}=&-\sum\limits_j [t_1(c^\dagger_{j, 1} c_{j, 2} + h.c.) + t_2(c^\dagger_{j, 2} c_{j+1, 1} + h.c.)] \notag\\
&+ \frac{\gamma}{2}\sum\limits_j (c^\dagger_{j, 2} c_{j, 1} - c^\dagger_{j, 1}c_{j, 2}), \label{eq:nHSSH}
\end{align}
where $c^{(\dagger)}_{j, \alpha}$ is the fermionic annihilation (creation) operator at site $j$ for the species or sublattices $\alpha=1, 2$. 
$t_1$ and $t_2$ are the usual hopping terms, while $\gamma$ is a dissipative chirality-preserving contribution to hopping. 
This model possesses the standard time-reversal, particle-hole and chiral symmetry, represented by
\begin{align}
\Kappa h_k \Kappa=h_{-k},\  \Kappa \sigma^z h_k \sigma^z \Kappa=-h_{-k},\ \{\sigma^z, h_k \} = 0,
\end{align}
where $\Kappa$ is the complex conjugation, $h_k$ is the Bloch Hamiltonian and $\sigma^z$ acts on the sublattice degree of freedom.
For $\gamma=0$, the model corresponds to the celebrated SSH model, which presents a topological phase for $\lvert t_1 \rvert<\lvert t_2 \rvert$ characterized by zero-energy edge states.
Its phase diagram, using the criterium of energy gap closing, can be analytically computed both with open (OBC) and periodic (PBC) boundary conditions \citep{Kunst2018}, and is shown in Fig.~\ref{fig:PD}\hyperref[fig:PD]{(a)}.
For PBC, it presents four different phases: two topological phases with non-trivial winding numbers, and two trivial phases, one of each directly connected to the two phases of the Hermitian SSH model.
For OBC on a finite system, there is no gap closing separating the two topological phases, and the boundaries separating the topological from the trivial phases differ from the periodic case.
The topological phases are characterized by zero-energy edge states.
As immediately visible in Fig.~\ref{fig:PD}\hyperref[fig:PD]{(b-d)}, there is no direct correspondence between the periodic bulk physics and the open edge physics. 
Gap closing points depend on the boundary conditions.
Interestingly, computation of the energy spectrum directly in the thermodynamic limit on a semi-infinite chain (see Appendix \ref{App-nH-SSH-SVD}) recovers the bulk phase diagram.
The non-Hermitian topological phase is characterized by an infinite-dimensional zero-energy Jordan block.
We therefore conclude that the phase diagram of the infinite open system differs from the infinite limit of the phase diagram of the open system.
Similar differences between finite and semi-infinite systems were observed in a study of Toeplitz matrices and operators\citep{Reichel1992}.

\subsection{Stability of edge modes and energies}

We now turn to the stability of the edge states observed in the non-Hermitian topological phase with OBC.
It was observed in Ref.~\onlinecite{Kunst2018} that these edge states are unstable to perturbations exponentially small in the total system size.
Indeed, a simple weak link $\tilde{t}_2\ll \lvert t_2-t_1\rvert$  connecting the two edges of the nH-SSH model can lead to a change of order $\lvert t_2-t_1 \rvert$ in the energy of the edge states, as illustrated in Fig.~\ref{fig:PD}\hyperref[fig:PD]{(e-f)}.
In the Hermitian limit, such perturbation only leads to a splitting of order $\tilde{t}_2$.
Similar results are obtained with a domain-wall configuration: connecting the two edges by a strongly gapped segment of the SSH model can lead to the disappearance of the edge state (see Appendix \ref{App-nH-SSH-sensitivity})
In contrast to what happens at the interface between different Hermitian topological phases, the two topologically distinct segments are not interfaced by edge states.
This is a direct consequence of the following inequalities.
For Hermitian matrices $A$ and $B$, the Weyl inequalities guarantee that the variation of eigenvalues due to perturbations are well-behaved, 
\begin{equation}
 \lvert E_j(A+B)-E_j(A) \rvert_\infty \leq \lvert \lvert B\rvert\rvert\label{eq:Weyl}
\end{equation}
where $E_j(A)$ (resp. $E_j(A+B)$) is the $j^\text{th}$ sorted eigenvalue of $A$ (resp. $A+B$) and the norm $\lvert\lvert\cdot\rvert\rvert$ is the spectral norm, that is to say the largest singular value of $B$ (largest absolute eigenvalue).
On the other hand, for $n \times n$ non-Hermitian matrices $A$ and $B$, the following inequality\citep{Krause1994} holds: 
\begin{equation}
d[E(A+B), E(A)]\leq c(n) (2 M)^{1-\frac{1}{n}}\lvert \lvert B\rvert \rvert^{\frac{1}{n}}, \label{eq:Instability}
\end{equation}
where $c(n)=\frac{16}{3\sqrt{3}} 2^{-\frac{1}{n}} <4$, $M=\max (\lvert \lvert A+B\rvert \rvert, \lvert \lvert A\rvert\rvert)$ and $d$ is the optimal matching distance:
\begin{equation}
d[E(A), E(B)]=\min\limits_{\pi \in S_n} \max\limits_j \ \lvert E_j(A)-E_{\pi(j)}(B) \rvert.
\end{equation}
$S_n$ is the group of all permutations.
Physically, for a Hermitian system, a perturbation of energy smaller than $\varepsilon$, cannot change the system's energy by more than $\varepsilon$, while an exponentially small perturbation of a non-Hermitian system can lead to macroscopic changes.
The aforementioned sensitivity to boundary conditions is the simplest example in which a physical perturbation leads to this exponential break-down of the stability of the eigenvalues.
Another concrete consequence of Eq. \eqref{eq:Instability} is the numerical noise often observed when diagonalizing non-Hermitian Hamiltonians, and the precision loss of standard linear algebra algorithms.

%

Such sensitivity of energy eigenvalues to perturbations calls into question the use of the winding of energies around special points as a topological invariant for non-Hermitian systems.
Working with a translation-invariant system with a finite number of bands, i.e., dealing with an effectively lower dimensional space, keeps under control the stability issue. 
Yet, instabilities immediately reappear when considering translation breaking perturbations, adding additional trivial bands, or considering interactions.
Additionally, the sensitivity to perturbations potentially jeopardizes the validity of all the usual approximations assumed in condensed matter systems: longer range tunneling or interactions that quickly decay may still have macroscopic effects.

\subsection{Singular values: solving both stability and bulk-boundary correspondence}

Instead of considering eigenvalues, we can in the same way study non-Hermitian systems using the singular value decomposition.
The SVD always satisfies the Weyl inequalities in Eq.~\eqref{eq:Weyl} and is thus \emph{well-behaved in the thermodynamic limit}.
Phase transitions are marked by a gap closing, i.e., a continuum of singular values reaching $0$.
The singular spectrum for the nH-SSH model can be obtained analytically (see Appendix \ref{App-nH-SSH-SVD}).
In the Hermitian limit, it is doubly degenerate, due to the particle-hole symmetry. 
For nonzero $\gamma$, half of the singular spectrum corresponds to the spectrum of the Hermitian SSH model with renormalized hopping $\tilde{t}_1=t_1+\frac{\gamma}{2}$ (without particle-hole induced degeneracy) and the other half to the Hermitian spectrum with $\tilde{t}_1=t_1-\frac{\gamma}{2}$.
As illustrated in Fig.~\ref{fig:PD}\href{fig:PD}{(a-d)}, the SVD phase diagram is identical for both OBC and PBC, and corresponds to the bulk energy phase diagram.
The non-Hermitian topological phase is now characterized by a \emph{single} zero-energy singular value, indicating the existence of a single zero-energy Jordan block (in contrast to two one-dimensional blocks in the Hermitian topological phase).
Similarly, we can study the stability of the zero singular modes to perturbations.
As expected, we recover the normal stability of Hermitian energy modes in Fig.~\ref{fig:PD}\href{fig:PD}{(e-f)}.
Domain walls between topologically distinct phases also translate into zero singular modes at the interfaces (see Appendix \ref{App-nH-SSH-sensitivity}).

\section{Topology through singular value decomposition}
\subsection{Topology in non-Hermitian systems}
Given the above observations, it is natural to shift language and reinterpret topology in terms of the SVD. 
By analogy with the Hermitian case, the actual value of the singular values should not matter, as long as they are nonzero.
The natural object to consider is therefore the unitary matrix $Q=UV^\dagger$, as a generalized flattened singular decomposition.
For a Hermitian system, $Q=P_+-P_-=2P_+-\mathds{1}=\mathds{1}-2P_-$, where $P_\pm$ are the projectors on the positive or negative energy bands.
$Q$ is then also Hermitian and satisfies $Q^2=\mathds{1}$.
The usual Hermitian topological invariants such as winding or Chern numbers can be rephrased in terms of $P_\pm$ and therefore also in terms of $Q$\citep{Bernevig2015}.
In the non-Hermitian case, there is no longer a simple notion of occupied and unoccupied bands, and the eigenvalues of $Q$ are no longer limited to be $\pm 1$ but can take any value in $\mathbb{U}(1)$. 
Nevertheless, the notion of bands remains, and both topological invariants and classification can be achieved.
Symmetries, such as particle-hole, time-reversal or chiral symmetry, play a similar role as in the Hermitian case.
Moreover, the symmetries of $H$ are also symmetries of $Q$, since $H=Q \sqrt{H^\dagger H}$, with $\sqrt{H^\dagger H}$ taken to be positive definite\citep{Gong2018}.

\subsection{Topological invariants}
\begin{figure}
\includegraphics[width=\linewidth]{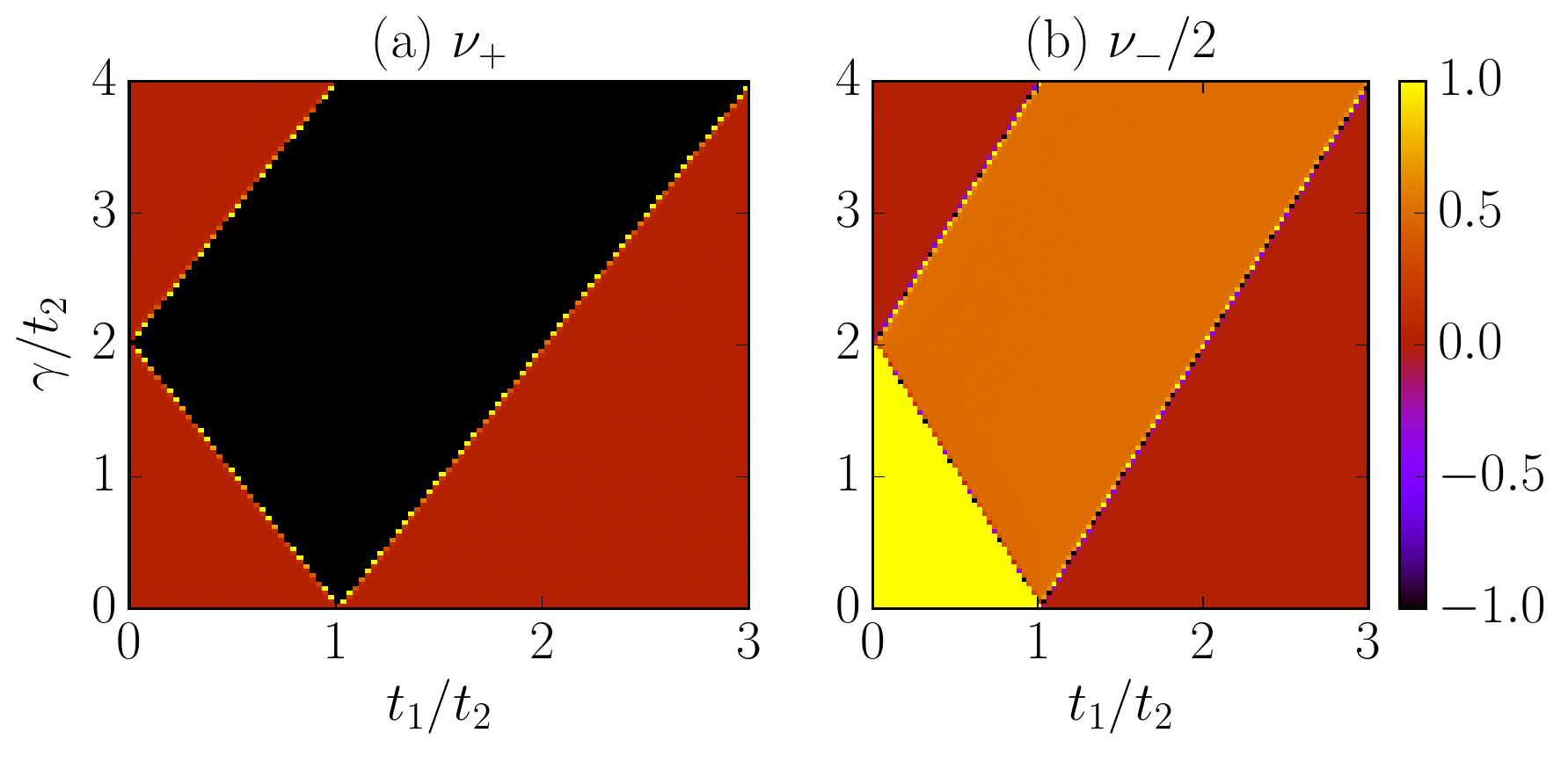}
\caption{Winding numbers computed for the non-Hermitian SSH model.  (a) $\nu_+$ is the total winding, and is trivial for a Hermitian Hamiltonian. (b) $\frac{\nu_-}{2}$ (the factor of $2$ is for convenience of representation) is the usual Hermitian winding number. We recover the PBC phase diagram and the $\mathbb{Z}\oplus\mathbb{Z}$ classification. The windings have been computed numerically for a periodic wire with $L=100$ unit-cells.}
\label{fig:WindingApp}
\end{figure}

In non\-interacting translation invariant systems, $Q$ can be written as the sum of $Q_{\vec{k}}$, where $Q_{\vec{k}}$ is obtained from the SVD of the Bloch Hamiltonian at momentum $\vec{k}$.
Topological classification of $Q$ then simply corresponds to the classification of the mappings $\vec{k}\rightarrow Q_{\vec{k}}$.
We now present a few examples in one and two-dimensions for different symmetry classes.
For pedagogical purposes, we focus on \emph{two-band models} (i.e., $Q_{\vec{k}}$ is a $2\times 2$ matrix), the generalization being generally straightforward.
For the non-Hermitian SSH model, the chiral symmetry implies 
\begin{equation}
Q_{\vec{k}} = \begin{pmatrix}
0 & q_1(\vec{k}) \\
q_2(\vec{k})& 0
\end{pmatrix}, 
\end{equation}
where $q_1$ and $q_2$ satisfy 
\begin{equation}
\det Q_{\vec{k}}=- q_1 q_2=\frac{\det H_{\vec{k}}}{\lvert\det H_{\vec{k}}\rvert}.
\end{equation}
The first homotopy group of the unitary matrices is $\mathbb{Z}$.
The homotopy group for chiral-symmetric matrices is therefore  $\mathbb{Z}\oplus \mathbb{Z}$.
The topological invariants associated to such a decomposition are:
\begin{equation}
\nu_+=\frac{i}{2 \pi} \int\limits_{\text{BZ}} \mathrm{Tr}(Q_{k}^\dagger \partial_k Q_{k}), \ \nu_-=\frac{i}{2 \pi} \int\limits_{\text{BZ}} \mathrm{Tr} (\sigma^z Q_{k}^\dagger \partial_k Q_{k}),
\end{equation}
or equivalently:
\begin{equation}
\nu_1= \frac{\nu_+-\nu_-}{2}\text{ and }\nu_2= \frac{\nu_++\nu_-}{2}.
\end{equation}
These two topological invariants match those previously introduced in Ref.~\onlinecite{Jiang2018, Yin2018} to describe the nH-SSH model, while the definitions are different.
Note that in the Hermitian case, $q_1=q_2^*$, and therefore $\nu_+$ trivially vanishes while $\nu_-$ is (twice) the usual winding number.
We illustrate the validity of these winding numbers by computing them over the full phase diagram of the nH-SSH model introduced in Eq. \ref{eq:nHSSH} in Fig.~\ref{fig:WindingApp}.
The obtained phase diagram matches the one obtained using an energy gap closing criterium with periodic boundary conditions.\\

We similarly define the Chern number from $Q$ in two dimensions.
Though there is no direct simple link between the Berry curvature and $Q$, one can write the Chern number as a winding of Wilson loops.
We define the Wilson loop operator by:
\begin{equation}
W_n(k_x)=\ln \left[ \mathrm{Tr} \prod\limits_{k_y \in BZ} U_{k_x, k_y} \Ket{n}\Bra{n} V^\dagger_{k_x, k_y}\right], \label{eq:WilsonLoop}
\end{equation}
where $\Ket{n}\Bra{n}$ is the projector on the $n$-$\mathrm{th}$ singular band (degeneracy in the Hermitian case can be taken care of by shifting by the identity matrix).
The Chern number is then simply given by the winding of $W_n$:
\begin{equation}
C_n=\frac{1}{2\pi} \oint dk_x \partial_{k_x} W_n(k_x).
\end{equation}
We have checked the validity of this definition in different non-Hermitian Chern insulator models (see Appendix \ref{2D}).

\subsection{Entanglement spectrum}

\begin{figure}
\includegraphics[width=\columnwidth]{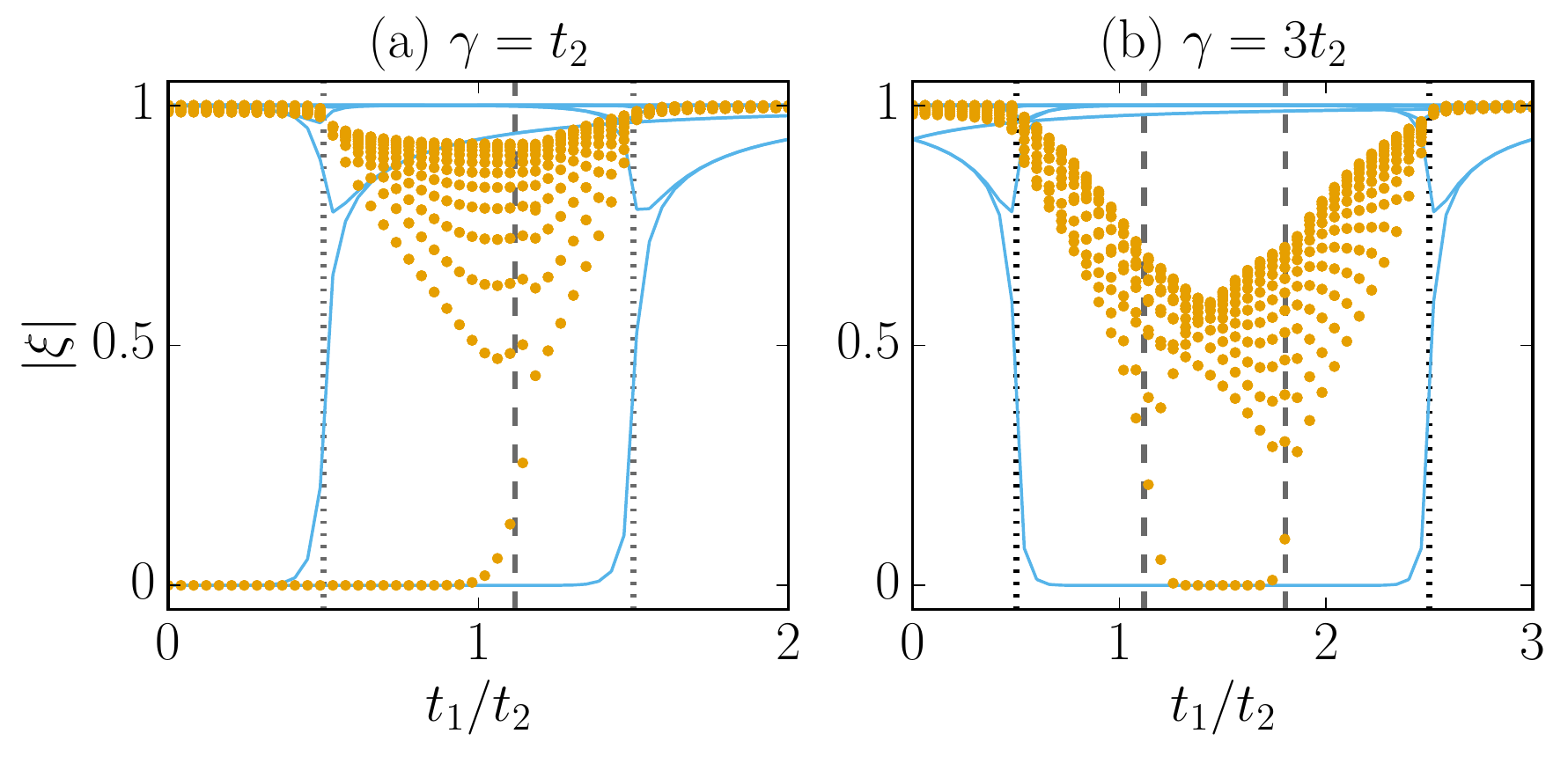}
\caption{Singular values (blue lines) and absolute value of the eigenspectrum (orange dots) of $Q_\mathcal{A}$ for the non-Hermitian SSH model. 
We consider a periodic wire with $L=200$ cells and consider a subsystem with $l=25$ unit-cells. 
The dashed (resp. dotted) vertical lines mark the OBC (resp. PBC) energy phase transitions. 
Both the singular and energy entanglement spectra match the behavior of the open system.}
\label{fig:entanglementSpectrum1D}
\end{figure}

A striking signature of the bulk-boundary correspondence in Hermitian topological system is found in the entanglement spectrum\citep{Li2008, Pollmann2010, Chandran2011, Qi2012}.
For Hermitian systems, the entanglement Hamiltonian is defined as $H_\mathrm{ent}=-\log \rho_\mathcal{A}$, where $\rho_\mathcal{A}$ is the reduced density matrix of the groundstate in the subsystem $\mathcal{A}$---its spectrum is the entanglement spectrum.
In a topological closed system, $H_\mathrm{ent}$ contains a universal low-energy part that corresponds to the edge theory of $H$ for OBC.
In a non\-interacting system, the entanglement spectrum can be obtained directly from the eigenvalues of the correlation matrix $C_\mathcal{A}=\Braket{c^\dagger_{\vec{r}} c_{\vec{r}'}}$ where $\vec{r}, \vec{r}'$ are restricted to $\mathcal{A}$\citep{Peschel2003}.
This matrix can be written as
\begin{equation}
C_\mathcal{A}=\frac{\mathds{1}-Q_\mathcal{A}^T}{2},
\end{equation}
where $Q_\mathcal{A}$ is the restriction of $Q$ to $\mathcal{A}$.
We therefore consider the  eigen- and singular-values of the matrix $Q_\mathcal{A}$ as analogues of the entanglement spectrum (up to some rescaling)\footnote{A shift by the identity of a non-Hermitian Hamiltonian is not trivial for its flattened representation $Q$}.
Topological zero modes of the Hamiltonian should translate into zero modes of $Q_\mathcal{A}$.
In Fig.~\ref{fig:entanglementSpectrum1D}, we show both the singular and the absolute eigenvalue  spectra of $Q_\mathcal{A}$ for the nH-SSH model of size $L$, computed from the periodic system for a segment $\mathcal{A}$  of length $l$.
The singular spectrum exactly matches the one obtained with an open boundary: the Hermitian topological phase is associated to two zero singular values, while the non-Hermitian one has only a single zero singular value.
Conversely, the energy spectrum exactly matches the one obtained for OBC, as long as $\mathcal{A}$ is not exactly half the system.
In other words, by considering the \emph{singular flattened Hamiltonian}, we have a true bulk-boundary correspondence: the physics of the open system is perfectly recovered for the periodic bulk Hamiltonian.

\begin{figure}[ht]
\includegraphics[width=\columnwidth]{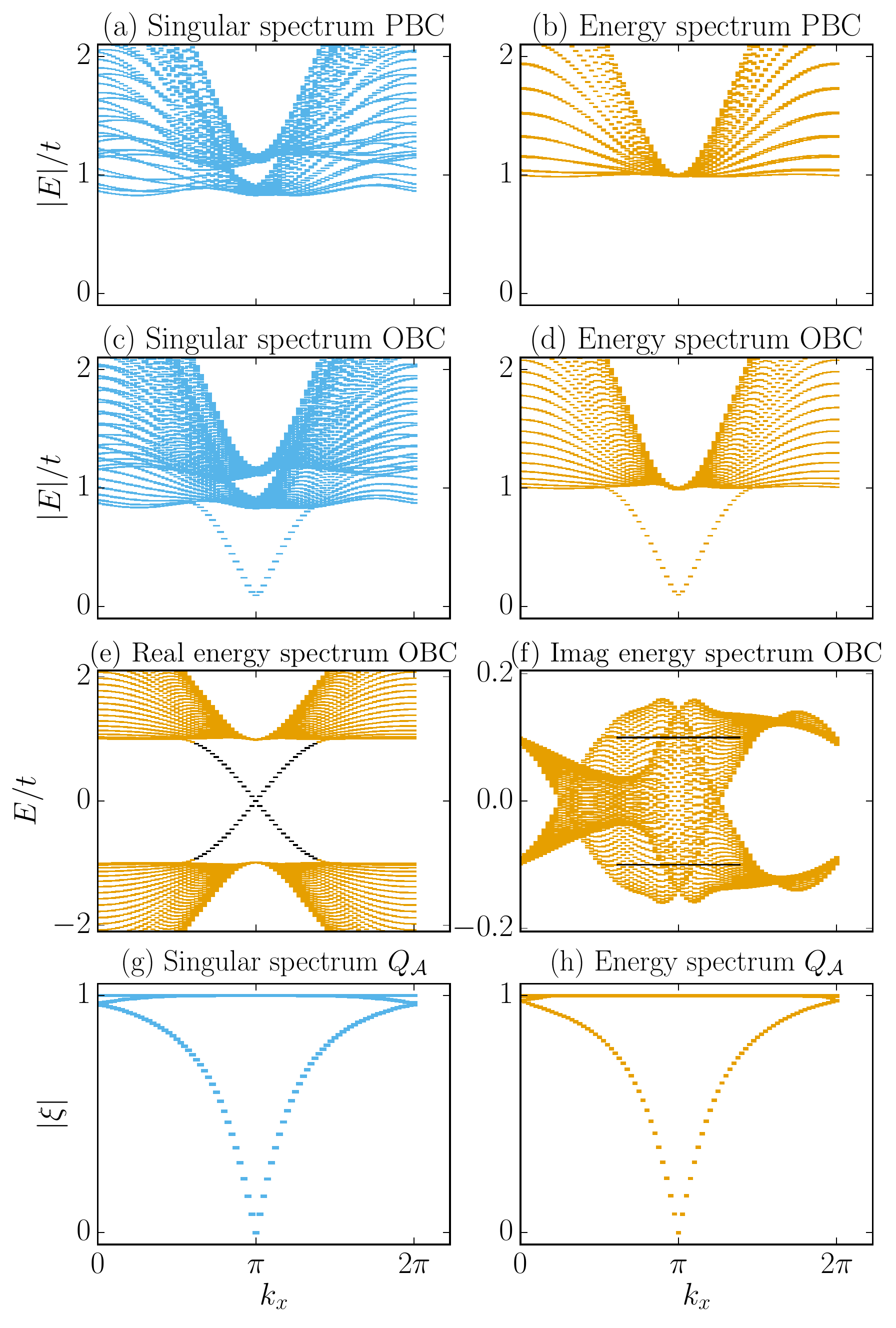}
\caption{(a) Singular and (b) absolute energy spectra for the non-Hermitian Chern insulator in its topological phase with $\gamma_x=\gamma_y=\delta \mu=0.1$ and $\mu=-t$ with PBC (see Appendix \ref{2D-Chern}). Both spectra are gapped. (c) Singular and (d) absolute energy spectra with OBC in the $y$ direction: chiral edge modes appear. (e-f) We represent the real and imaginary part of the previous energy spectrum. For clarity, the chiral modes are in black. The imaginary part of the energy does not cancel: the modes are dissipative. (g-h) The spectra of the corresponding $Q_\mathcal{A}$ also show a chiral boundary mode, as expected, but without a dissipative component. We considered a strip of width $l_y=30$.}
\label{fig:entSpectrum2D}
\end{figure}

Similar results can be obtained for the two-dimensional model Chern insulator introduced in Ref.~\onlinecite{Shen2018} (see Appendix \ref{2D-Chern} for the microscopic model).
They are summarized in Fig. \ref{fig:entSpectrum2D}.
We consider a periodic system of length $L_x=L_y=80$ unit-cells, and our subsystem $\mathcal{A}$ is a strip of length $l_x=L_x$ and $l_y=30$ of the torus.
The spectra of $Q_\mathcal{A}$ indeed shows the presence of the chiral modes even in the presence of the non-Hermitian dissipation, with an interesting caveat.
In the open system, the chiral energy modes always have a finite dissipative part, i.e., the imaginary part of the energy does not vanish, with an effective low-energy dispersion relation given by:
\begin{equation}
E_\pm(k_x)=\pm\left[v_F (k_x-\pi) + i e_0\right],
\end{equation}
where $v_F$ is the Fermi velocity and $e_0$ a constant. 
Correspondingly, the singular modes do not exactly reach zero. 
Conversely, the spectra of $Q_\mathcal{A}$ present clear low-energy modes which do vanish at $k_x = \pi$, both for energies and singular values. 
We find that the effective low-energy modes in the spectrum of $Q_\mathcal{A}$ are well fitted by:
\begin{equation}
\xi_\pm(k_x)=\pm\left[v_F (k_x-\pi) + i h_0 (k_x-\pi)^2\right],
\end{equation}
where $h_0$ is a constant. 
The dissipative nature of the chiral boundary modes is therefore not fully captured by the entanglement spectrum here. 
Similar discrepancies between boundary modes and entanglement spectrum have also been observed in Hermitian systems\citep{Turner2010}.
Note that in this system, there are significant differences between singular and eigenspectra, even in the bulk. 
The zero singular modes of $Q_\mathcal{A}$ are also a sign of the non-trivial topology of the model, even though the spectrum is technically always gapped.
This hidden topological structure in our analog of the entanglement spectrum is in partial contradiction with the classification of Ref. \onlinecite{Gong2018}.
While the Hamiltonian itself does not have topological \emph{zero} modes, it has non-trivial topological properties.
Additional results on the model introduced in Ref. \citep{Kunst2018} can be found in Appendix \ref{2D-Flore}.
This model breaks bulk-boundary correspondence for the energies.
Both the open system and the spectrum of $Q_{\mathcal{A}}$ shows the presence of a flat singular zero bands.

\section{Conclusions and discussions}

In this work, we have shown how the bulk-boundary correspondence, considered to be broken in non-Hermitian systems, can actually be restored by shifting from eigenvalue considerations to a singular value decomposition.
Due to the instability of the eigenspectrum to small perturbations, a general topological classification of non-Hermitian models in terms of energies is not formally well-defined.
Our framework provides a path towards a natural classification of Hamiltonians in terms of their flattened singular decomposition $Q=UV^\dagger$.
We discussed how to recover topological invariants from $Q$ and gave some concrete examples of topological non-Hermitian phases.
These topological invariants have the advantage to be explicit functions of the flattened Hamiltonian, and therefore one can directly use the well-known real-space reformulations of the topological invariants\citep{Niu1985, Prodan2010, Mondragon2014} for systems which are not translation invariants (e.g. in the presence of disorder).
Finally, we proposed an analog of the entanglement spectrum to non-Hermitian systems.
We showed that indeed, the bulk system contains complete information on the edge physics of both eigenvalues and singular values.
Note that this approach for topology in non-Hermitian systems partially coincides with the one introduced in Ref.~\onlinecite{Gong2018}, while coming from a completely different perspective.
The mapping to an effective chiral model can be directly applied to the unflattened Hermitian Hamiltonian: $\tilde{H}=H \otimes \tau^+ + h.c.$, where $2\tau^+=\tau^x+i\tau^y$ are Pauli matrices acting on an additional degree of freedom.
Its eigenvalues are $\pm \Lambda$, with $\Lambda$  the singular values of $H$\citep{Porras2018}.
The topological zero modes of $\tilde{H}$ (or equivalently $\tilde{Q}=Q \otimes \tau^+ + h.c.$), corresponding to the different topological classes, are nothing but the zero singular values of $H$.
Our results are immediately applicable to the classifications based on a single forbidden energy\citep{Gong2018, Kawabata2018, Zhou2018}.
A full generalization to topology based on forbidden manifold\citep{Kawabata2018} is still an open problem.
It was also shown in Ref. \onlinecite{Porras2018} that the singular values can be linked to the steady-state coherences in the Lindbladian evolution of bosonic systems.
Our discussion can be straightforwardly extended to multiband problems, generally replacing the usual notion of energy band-gap by singular-value band-gap.
The spectrum of $Q_\mathcal{A}$ can still be used as an analog to the entanglement spectrum.
Extending this entanglement spectrum analog to each singular band independently is left to future work.
Generalization of this approach to Lindblad systems would be an interesting subject of research.
Similarly, comparison of these results to systems with a complete Hamiltonian description of the dissipative part would answer important issues.
Indeed, non-Hermitian models are nothing but a simple approximation of complex Hermitian systems: whether this classification would carry through is an open question.
There, the instability of non-Hermitian systems to small perturbations could be addressed directly.

\begin{acknowledgments}
This work was supported by the ERC Starting Grant No. 679722 and the Knut and Alice Wallenberg Foundation 2013-0093. We thank Thom\'{a}\v{s} Bzdu{\v{s}}ek, Adrien Bouhon, Flore Kunst, Simon Lieu, Diego Porras and Henning Schomerus for useful discussions.
\end{acknowledgments}

\appendix

In these Appendices, we provide further technical details and additional examples of topological invariants and entanglement spectrum in one and two dimensions.
Specifically, we provide a complete solution of the chiral non-Hermitian Su-Schrieffer-Heeger model. We compute the singular values for periodic, open and semi-infinite systems.
We discuss the stability of topological edge states in the presence of domain walls.
We obtain numerically the phase diagram from the computation of the two winding numbers introduced in the main text and we discuss in more details the entanglement spectrum.
Secondly, we provide two examples of two-dimensional topological non-Hermitian models.
We compute the relevant Chern number from its Wilson loop formulation, and show that the analogue of the entanglement spectrum introduced in the main text is still valid.

\section{The chiral nH-SSH model}\label{App-nH-SSH}
\subsection{Phase diagram and boundary conditions}\label{App-nH-SSH-PD}
The non-Hermitian chiral SSH model is a paradigmatic example of a chiral-symmetric topological model in one-dimension.
Its real space Hamiltonian is given by Eq. 2 of the main text, and the corresponding momentum Hamiltonian is:
\begin{equation}
H_\mathrm{nH-SSH}=\sum\limits_k \Psi_k^\dagger [\vec{n}(k).\vec{\sigma} + i \vec{d}(k).\vec{\sigma}] \Psi_k, 
\end{equation}
with $\Psi^\dagger_k=(c^\dagger_{k, 1}, c^\dagger_{k, 2})$, $\vec{n}(k)=(-t_1-t_2 \cos k, -t_2 \sin k, 0)$ and $\vec{d}(k)=(0, \frac{\gamma}{2}, 0)$.
The energies at momentum $k$ are given by:
\begin{equation}
E_k^2 = t_1^2 + t_2^2 + 2t_1 t_2 \cos k - \frac{\gamma^2}{4} - i\gamma t_2 \sin k.
\end{equation}
The gap-closing conditions for this two-band non-Hermitian model are
\begin{equation}
\lvert\lvert\vec{n}(k)\rvert\rvert=\lvert\lvert\vec{d}(k)\rvert\rvert \text{ and } \vec{n}(k).\vec{d}(k)=0,
\end{equation}
which straightforwardly lead to phase boundaries for a periodic system given by $t_1\pm t_2 = \pm \frac{\gamma}{2}$.
Conversely, for a finite open system, the phase boundaries were computed analytically in Ref.~\onlinecite{Kunst2018} and correspond to $(t_1\pm t_2)^2 = \frac{\gamma^2}{4}$. The mismatch between the two is illustrated in Fig. 1 in the main text.

Let us now consider the semi-infinite limit. 
We take a lattice defined for $j\geq 0$ and look for zero-energy left- and right- eigenstates of the form:
\begin{equation}
\Ket{\psi_0^R}=\sum\limits_{j\geq 0} r_j c^\dagger_{j, 1}\Ket{0} \text{ and } \Ket{\psi_0^L}=\sum\limits_{j\geq 0} l_j c^\dagger_{j, 1}\Ket{0}.
\end{equation}
Straightforward algebra leads to 
\begin{equation}
r_j\propto \left(-\frac{t_1+\gamma/2}{t_2}\right)^j \text{ and } l_j\propto \left(-\frac{t_1-\gamma/2}{t_2}\right)^j.
\end{equation}
The normalization condition applied separately to each state leads to the PBC phase diagram, while the mutual normalization $\Braket{\psi_0^L \lvert \psi_0^R}=1$ leads to the OBC phase diagram. 
The latter condition is not necessarily correct if these two states are the extremal states of a Jordan block. Let us look for a state $\Ket{\psi_1^R}\propto\sum\limits_{j\geq 0} r_{j, 1} c^\dagger_{j, 2}\Ket{0}$ such that $H\Ket{\psi_1^R}=\ket{\psi_0^R}$. 
If such a state exists, then $H$ is not diagonalizable and the zero eigenvalue is associated to a non-trivial Jordan block. 
The coefficients in $\Ket{\psi_1^R}$ should satisfy the following set of equations:
\begin{align}
-(t_1-\frac{\gamma}{2}) r_{0, 1} &= 1 \\
-(t_1-\frac{\gamma}{2}) r_{m, 1} -t_2 r_{m-1, 1} &= \left(-\frac{t_1+\gamma/2}{t_2} \right)^m,
\end{align}
which admit the following solution:
\begin{equation}
r_{j, 1}= \left(-\frac{t_1+\gamma/2}{t_2} \right)^j \left(\frac{1}{\gamma/2-t_1} - j \frac{t_1+\gamma/2}{t_2^2} \right)
\end{equation}
if $t_2^2-t_1^2+\gamma^2/4=0$, and 
\begin{multline}
r_{j, 1}= \left(-\frac{t_1+\gamma/2}{t_2} \right)^j \frac{t_ 1+\gamma/2}{t_2^2-t_1^2+\gamma^2/4} + \\
\left(\frac{t_2}{\gamma/2-t_1} \right)^j \left( \frac{1}{\gamma/2-t_1}-\frac{t_ 1+\gamma/2}{t_2^2-t_1^2+\gamma^2/4} \right)
\end{multline}
otherwise. 
Indeed, the matrix admits a Jordan block for the $0$ eigenvalue, which imply that $\Braket{\psi_0^L | \psi_0^R}=1$ is not the proper normalization condition. 
Our construction actually does not stop with only one such state, i.e., the Jordan block is not just a $2\times 2$ block. 
Similar calculation leads to an infinite tower of states satisfying $H \Ket{\psi_{k+1}}=\Ket{\psi_k}$, with 
\begin{equation}
r_{m, 2k}=\sum\limits_{j=0}^k \alpha_k^e m^k (-\frac{t_1+\gamma/2}{t_2})^m + \sum\limits_{j=0}^{k-1} \beta_k^e m^k (\frac{t_2}{\gamma/2-t_1})^m, 
\end{equation}
\begin{equation}
r_{m, 2k+1}=\sum\limits_{j=0}^k \alpha_k^o m^k (-\frac{t_1+\gamma/2}{t_2})^m + \sum\limits_{j=0}^{k} \beta_k^o m^k (\frac{t_2}{\gamma/2-t_1})^m. 
\end{equation}
The $\alpha_k^{o, e}$ and $\beta_k^{o, e}$ coefficients can be systematically computed by recurrence.  
The zero-energy space is actually infinite in the thermodynamic limit. 
This prevents us from properly determining the phase boundary by this approach, though the form of the coefficients that appear tend to imply that the PBC phase diagram is the correct one. 
To clarify this picture, let us turn to the computation of the singular values.

\subsection{Evaluating the singular value decomposition} \label{App-nH-SSH-SVD}
We first evaluate the singular values of the matrix $h_k$. 
Computing singular values in general is a cumbersome task. 
Here we use the fact that they are the eigenvalues of $h^\dagger_k h_k$. We then directly obtain
\begin{equation}
\lambda_{\pm}(k)^2=(t_1+t_2 \cos k \pm \frac{\gamma}{2})^2 + t_ 2^2\sin^2 k.
\end{equation}
We see immediately that there is no simple link between eigenvalues and singular values, even in the case of a $2\times 2$ matrix. 
The singular values are zero when the eigenvalues of $h(k)$ are also zero, and we recover the PBC phase diagram.
We also see that $\gamma$ acts as a translation of the hopping parameter $t_1$. Let us prove this result for arbitrary boundary conditions. One can rewrite the nH-SSH model as:
\begin{equation}
H_\mathrm{nH-SSH}= -t_1 \sigma^x \otimes \text{Id} -t_2(\sigma^+ \otimes T^r +\sigma^-\otimes T^l)+ i \frac{\gamma}{2} \sigma^y \otimes \text{Id}
\end{equation}
where $\sigma$ acts on the pseudo-spin subspace and the other operators on the unit-cell subspace. We use the convention $2\sigma^\pm=\sigma^x\pm i \sigma^y$.
$T^r$ ($T^l$) is the translation operator to the right (left), taking into account the proper boundary conditions. We then obtain:
\begin{multline}
H_\mathrm{nH-SSH}^\dagger H_\mathrm{nH-SSH} = (t_1+\frac{\gamma}{2} \sigma^z)^2 + (t_1+\frac{\gamma}{2} \sigma^z)t_2 (T^l+T^r)  \\
+t_2^2 (\sigma^- \sigma^+ \otimes T^l T^r +\sigma^+ \sigma^- \otimes T^r T^l)
\end{multline}
This matrix is actually diagonal in the pseudo-spin space for all $\gamma$. 
For $\gamma=0$, the two pseudo-spin flavors have the same eigenvalues, and therefore lead to a double degeneracy of the singular spectrum, which is nothing but the $\pm E$ particle-hole symmetry of the Hermitian model. 
On the other hand, for $\gamma\neq 0$, the two flavors correspond to two Hermitian models with different effective $t_1$. 
Half the non-Hermitian singular spectrum therefore corresponds to the positive energy spectrum of a Hermitian SSH (H-SSH) model with $t_1^\text{eff}=t_1+\frac{\gamma}{2}$ (without the particle-hole degeneracy) and  half to the spectrum of a H-SSH model with $t_1^\text{eff}=t_1-\frac{\gamma}{2}$. 
This mapping immediately implies that the phase diagram given by the SVD with open or periodic boundary conditions or in the semi-infinite limit is identical to the phase diagram derived from the energies with PBC. 
It also tells us that the non-Hermitian topological phase corresponds to having only one copy of the SSH model in its topological phase, and therefore only a single zero singular value. 
Note that in the finite system, the two energy edge states are not necessarily in the same Jordan block, as they have an exponentially small residual energy. 

\subsection{Stability of the edge states} \label{App-nH-SSH-sensitivity}

\begin{figure}
\includegraphics[width=\columnwidth]{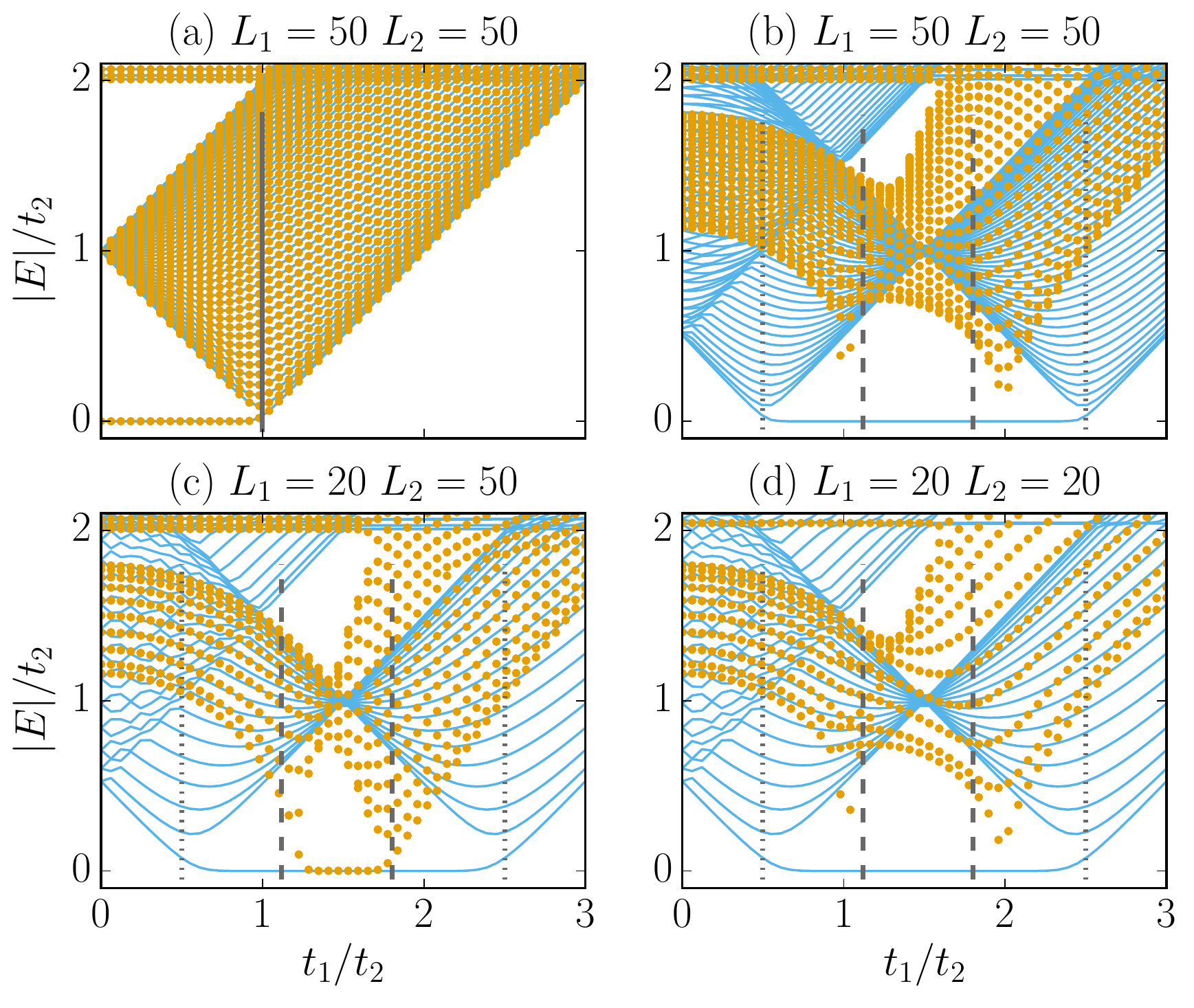}
\caption{Singular (blue lines) and absolute energy (orange dots) spectrum for different domain-wall configurations. We fix $T_1=4t_2$, $T_2=2t_2$. Vertical lines mark the phase transition of the energy spectrum for OBC (dashed) and PBC (dotted). Both are identical (full line) in the Hermitian case. (a) In the Hermitian limit $\gamma=0$, zero-energy states appear at the interfaces when the first half of the wire is in the topological phase. (b-d) In the non-Hermitian case ($\gamma=3t_2$), the exponentially small coupling induced by the trivial part can gap out the energy states while the zero singular state survives. Pardoxically, a smaller topological part is more resilient.}
\label{fig:DW}
\end{figure}

In the main text, we have studied the resilience of the edge states to the introduction of a direct coupling between the two edges. 
A fair critic could argue that directly coupling the edges should gap them out, albeit by a smaller amount. 
To show that this result is not merely the gapping out of the edge modes, we propose a slightly different scheme based on a periodic wire whose first half is in the non-Hermitian topological phase and second half is in the Hermitian trivial phase. We choose parameters such that the trivial part is strongly gapped.
This setup is a common setup for topological studies: at the interface between the trivial and topological phases, we expect the appearance of the topological phase's edge modes.\\
To be concrete, the Hamiltonian we study is given by Eq. 2 in the main text, for $L=L_1+L_2$ unit cells, but with the following site-dependent couplings:
\begin{align}
t_{1, j}&=\left\lbrace\begin{array}{cc} t_1 & \text{if }1\leq j \leq L_1 \\ T_1 & \text{otherwise}  \end{array} \right. \notag\\ 
\gamma_{j}&=\left\lbrace\begin{array}{cc} \gamma & \text{if }1\leq j \leq L_1 \\ 0 & \text{otherwise}  \end{array}\right.\\
t_{2, j, j+1}&=\left\lbrace\begin{array}{cc} t_2 & \text{if }1\leq j \leq L_1 \\ T_2 & \text{otherwise}  \end{array}\right.\notag\\ 
t_{2, 1, L}&=t_2\notag
\end{align}
The results are given in Fig. \ref{fig:DW}. We fixed $T_1=4t_2=2T_2$ such that the second part of the wire is deep in the trivial phase with a gap of order $2 t_2$, much larger than the gap of the first half. The exponentially small coupling induced by the trivial part is enough to gap out the energy zero states if $L_1/L_2$ is large enough. The zero singular value that appears is itself unaffected and behaves as expected in a topological system.

\subsection{Entanglement spectrum} \label{App-nH-SSH-WN}

\begin{figure}
\includegraphics[width=\linewidth]{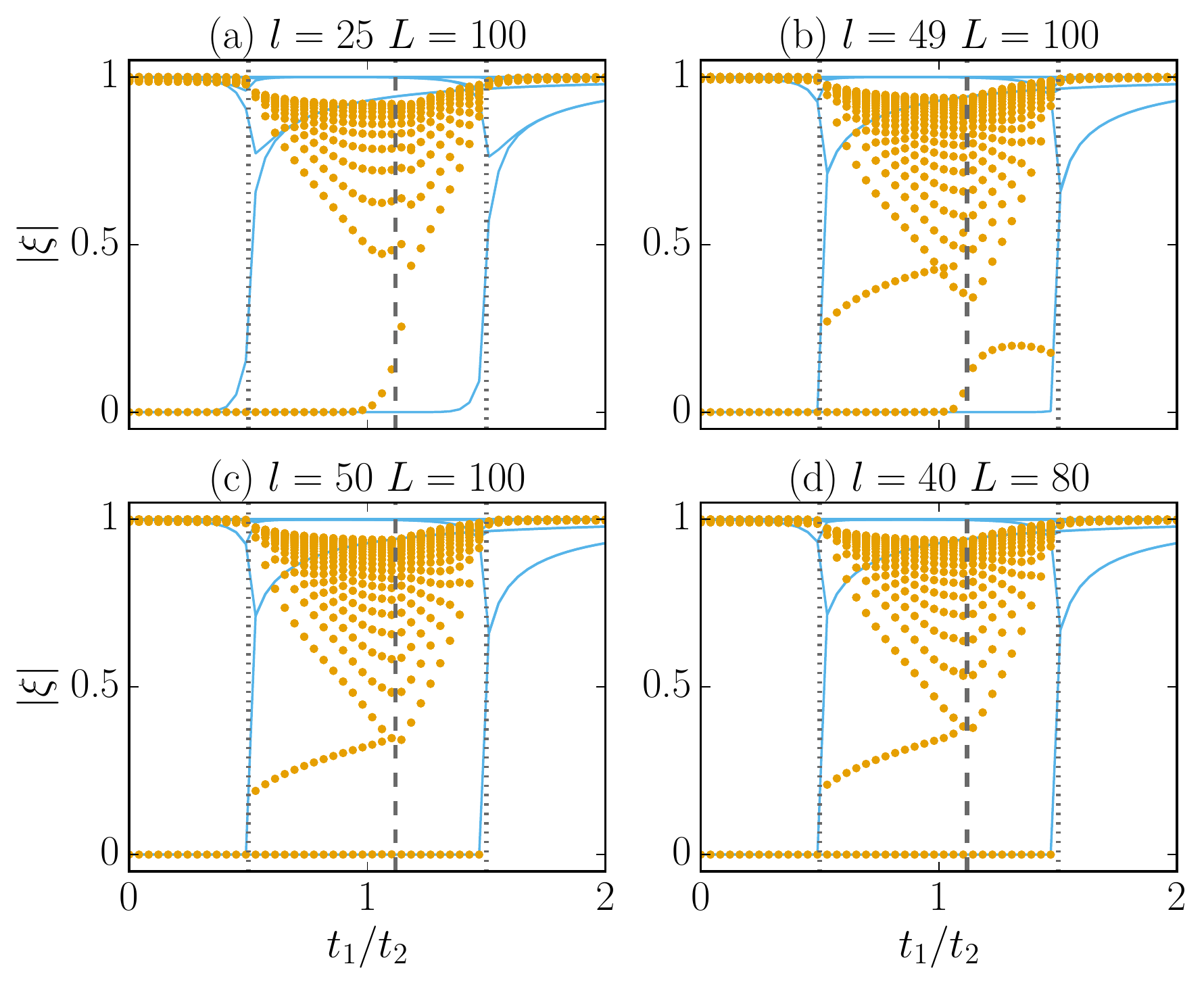}
\caption{Singular values (blue lines) and absolute value of the eigenspectrum (orange dots) of $Q_\mathcal{A}$ for the non-Hermitian SSH model for $\gamma=t_2$. 
We consider a  subsystem with $l$ unit-cells for a total wire of length $L$. The dashed lines mark the OBC energy phase transitions while the dotted lines mark the PBC energy and singular phase transitions. Except when $L=2l$, both the singular and energy spectra match the behavior of the open system. When $L=2l$, it behaves here as in the semi-infinite limit.}
\label{fig:eS-size}
\end{figure}

Finally,  Fig.~\ref{fig:eS-size} represents the size dependence of the spectra of $Q_\mathcal{A}$, our analog of the entanglement spectrum. 
The half-system cut is anomalous and does not present the same physics as the others. 
While we do not have a full understanding  of this feature and of its universality, the absence of continuity could be explained by the instability of non-Hermitian systems.
Note that neither the determinant or the trace of $Q_\mathcal{A}$ or any of its relevant submatrices present this finite-size behavior.

\section{Two-dimensional models}\label{2D}

\subsection{Non-Hermitian Chern insulator}\label{2D-Chern}

As a two-dimensional example for both the computation of the Chern number and the entanglement spectrum, we study a non-Hermitian generalization of the two-band Chern insulator introduced in Ref.~\onlinecite{Shen2018}. This model still exhibits the usual bulk-boundary correspondence, but its proximity to the standard Hermitian Chern insulator make it an ideal benchmark for our methods. We parametrize the Bloch Hamiltonian $h(\vec{k})$ as:
\begin{equation}
 h(\vec{k})=\left[ \vec{n}(\vec{k}) + i \vec{d}(\vec{k})\right].\vec{\sigma},
 \end{equation} 
with
\begin{align}
\vec{n}(\vec{k})&=(\Delta_x \sin k_x, \Delta_y \sin k_y, - \mu - t \cos k_x -t \cos k_y)\\
\vec{d}(\vec{k})&=(\gamma_x , \gamma_y, \delta \mu).
\end{align}
If the two fermionic species are spin polarizations, $\mu$ corresponds to a Zeeman field, $t$ a hopping between lattice sites, $\Delta_x$ and $\Delta_y$ are spin orbit couplings, and $\gamma_x$ and $\gamma_y$ are constant dissipative spin-flip terms, while $\delta \mu$ is a local source or drain coupled to the spin polarization. 
In the following, for simplification, we take $t=\Delta_x=\Delta_y=1$.

In the Hermitian limit $\vec{d}=\vec{0}$, for non-zero $\Delta_x$ and $\Delta_y$, the system is in a trivial gapped phase for $\lvert \mu \rvert> 2\lvert t \rvert$, and in a gapped topological phase for $\lvert \mu \rvert< 2\lvert t \rvert$. Two distinct topological phases exist, separated by a gapless point at $\mu=0$. 
In both topological phases, each band is characterized by a Chern number $\pm 1$ and chiral edge states are present.
Both topological phases host chiral edge states.

\begin{figure}
\includegraphics[width=\columnwidth]{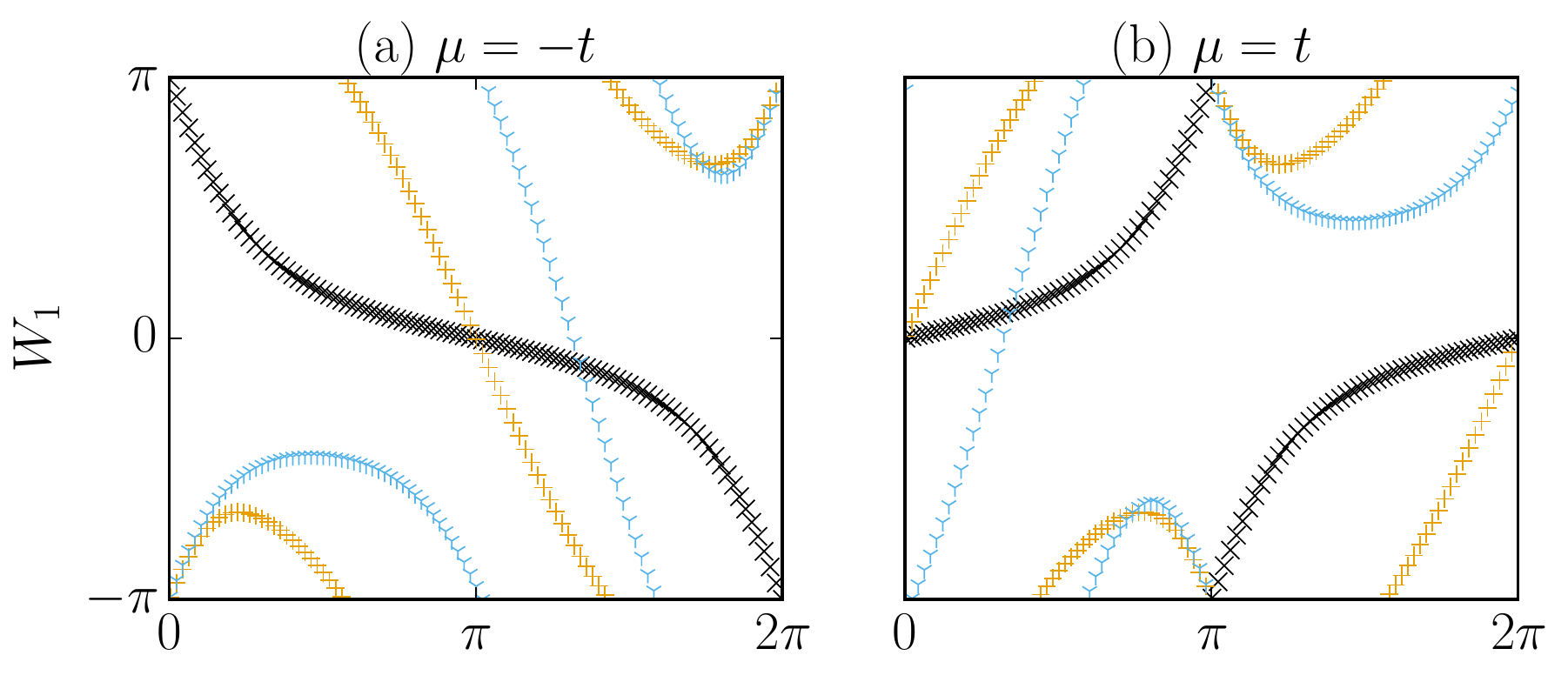}
\caption{Wilson loop computed for the non-Hermitian Chern insulator, focusing on the lowest singular band. For numerical convenience, we shift the Hamiltonian's diagonal entries by $0.1$ before computing the SVD. Black crosses correspond to the Hermitian limit $\gamma_x=\gamma_y=\delta \mu=0$, orange pluses to $\gamma_x=\gamma_y=0.1$, $\delta \mu=0$ and blue stars to $\gamma_x=\gamma_y=\delta \mu=0.1$. We considered a periodic system of $L_x=150$ by $L_y=60$ unit cells. The winding of the Wilson loops correctly capture the Chern number of the topological phase, even in the presence of non-Hermitian terms.}
\label{fig:WilsonLoop}
\end{figure}

Obtaining the complete phase diagram analytically in the presence of the non-Hermitian terms is a fairly involved computation, so we focus in this appendix on the phases adiabatically connected to the Hermitian phases. 
Using Weyl inequalities for singular values, it is straightforward to show that all three phases survive the presence of non-Hermitian perturbations smaller than their gap.
Figure ~\ref{fig:WilsonLoop} presents the winding number of the Wilson loops defined in Eq. 11 in the main text, for different sets of parameters. We have selected here the lowest singular band. The Chern number indeed survives the presence of non-Hermitian terms.

\subsection{Chern insulator with broken bulk-boundary correspondence}\label{2D-Flore}
In the previous example, both energy and singular values led to similar phase diagrams: there was no breakdown of the bulk-boundary correspondence. 
We also check that our approach is valid in a toy model introduced in Ref.~\onlinecite{Kunst2018}, where the broken bulk-boundary correspondence is restored in our analog of the entanglement spectrum.
Here the Bloch Hamiltonian is given by:
\begin{align}
n_x&=t_1+\delta \cos k_x + (t_1-\delta \cos k_x) \cos k_y,\nonumber \\
 n_y&=(t_1-\delta \cos k_x) \sin k_y,\ n_z=t_1-\Delta \sin k_x,  \nonumber \\
d_y&=\frac{\gamma}{2},\ d_x=d_z=0. \label{eq:2DFlore}
\end{align}

\begin{figure}[ht!]
\includegraphics[width=\columnwidth]{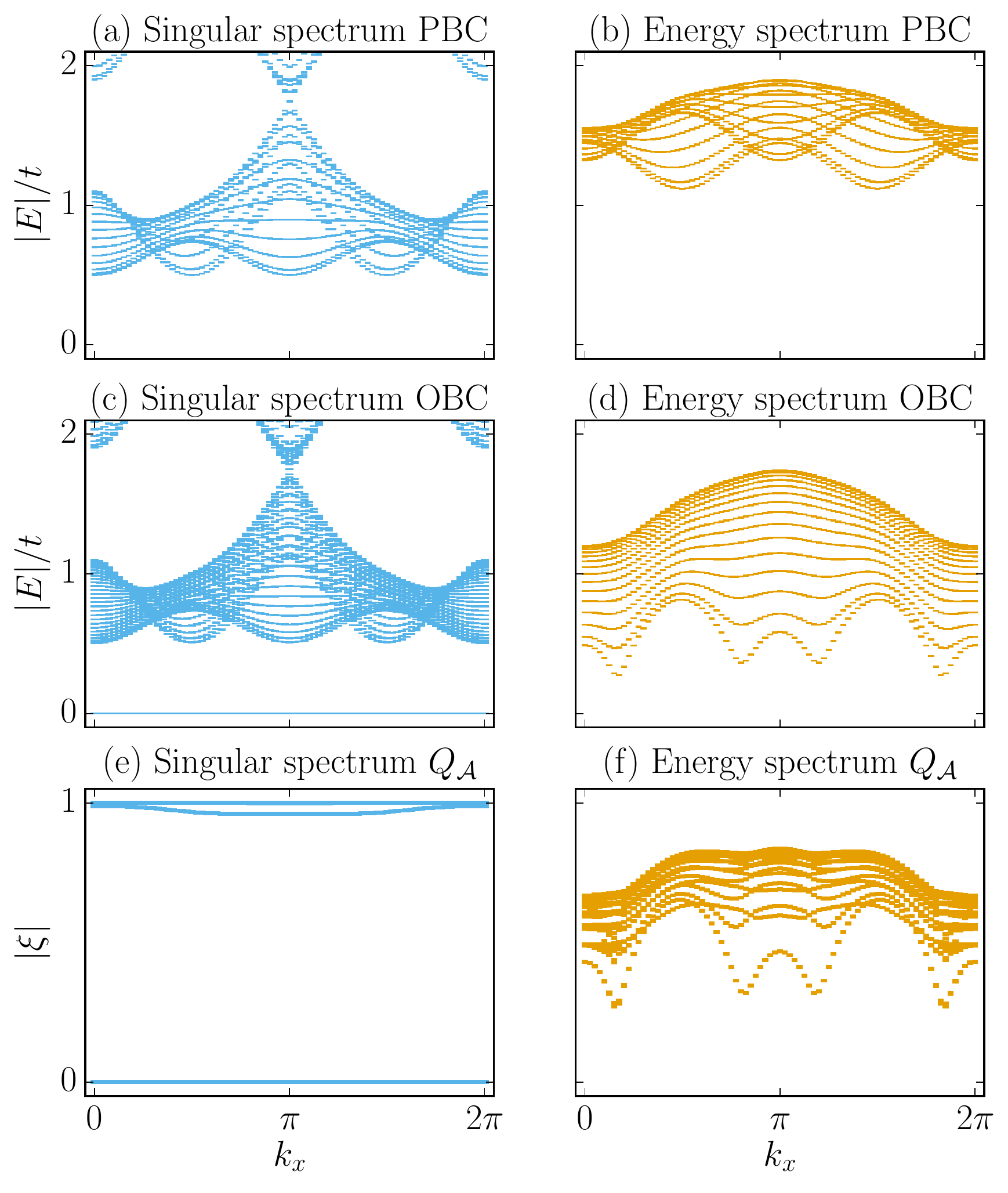}
\caption{(a) Singular and (b) absolute energy spectra for the model in Eq.~\eqref{eq:2DFlore} in its topological phase for PBC. Both spectra are gapped. We take $L_x=L_y=80$. (c) Singular and (d) absolute energy spectra for a system open in the $y$-direction: a flat zero-singular band appears, while the energy spectrum stays gapped. (e-f) The spectra of the corresponding $Q_\mathcal{A}$ matches the spectrum of the open system. We considered a strip of width $l_y=30$.
}
\label{fig:entSpectrum2D-Floremodel}
\end{figure}

For pedagogical purposes, we focus on a single point of the phase diagram: $t_1=\Delta$, $\delta=0.2t_1$ and $\gamma=3$. 
For PBC, the system is then gapped both for singular values and energies. 
If we now consider a system periodic in the $x$ direction and open in the $y$ direction, a zero-singular flat band appears and the energy spectrum remains gapped without zero modes. 
Computation of the entanglement spectrum for a strip of finite width in the $y$ direction for the periodic system exactly matches what we observe in the open system. 
Results are summarized in Fig.~\ref{fig:entSpectrum2D-Floremodel}. 
We can also compute the Chern number through the computation of the winding of the Wilson loops defined in Eq. (11) in the main text. 
We find that the Hamiltonian is nontrivial, with a Chern number $-1$ for the lowest singular band.
The presence of a flat-band of zero modes is typical in such two-dimensional anisotropic hopping model.
\vspace{0.3cm}

\vspace*{-0.25cm}
\bibliography{nonHermitian}

\begin{thebibliography}{56}%
\makeatletter
\providecommand \@ifxundefined [1]{%
 \@ifx{#1\undefined}
}%
\providecommand \@ifnum [1]{%
 \ifnum #1\expandafter \@firstoftwo
 \else \expandafter \@secondoftwo
 \fi
}%
\providecommand \@ifx [1]{%
 \ifx #1\expandafter \@firstoftwo
 \else \expandafter \@secondoftwo
 \fi
}%
\providecommand \natexlab [1]{#1}%
\providecommand \enquote  [1]{``#1''}%
\providecommand \bibnamefont  [1]{#1}%
\providecommand \bibfnamefont [1]{#1}%
\providecommand \citenamefont [1]{#1}%
\providecommand \href@noop [0]{\@secondoftwo}%
\providecommand \href [0]{\begingroup \@sanitize@url \@href}%
\providecommand \@href[1]{\@@startlink{#1}\@@href}%
\providecommand \@@href[1]{\endgroup#1\@@endlink}%
\providecommand \@sanitize@url [0]{\catcode `\\12\catcode `\$12\catcode
  `\&12\catcode `\#12\catcode `\^12\catcode `\_12\catcode `\%12\relax}%
\providecommand \@@startlink[1]{}%
\providecommand \@@endlink[0]{}%
\providecommand \url  [0]{\begingroup\@sanitize@url \@url }%
\providecommand \@url [1]{\endgroup\@href {#1}{\urlprefix }}%
\providecommand \urlprefix  [0]{URL }%
\providecommand \Eprint [0]{\href }%
\providecommand \doibase [0]{http://dx.doi.org/}%
\providecommand \selectlanguage [0]{\@gobble}%
\providecommand \bibinfo  [0]{\@secondoftwo}%
\providecommand \bibfield  [0]{\@secondoftwo}%
\providecommand \translation [1]{[#1]}%
\providecommand \BibitemOpen [0]{}%
\providecommand \bibitemStop [0]{}%
\providecommand \bibitemNoStop [0]{.\EOS\space}%
\providecommand \EOS [0]{\spacefactor3000\relax}%
\providecommand \BibitemShut  [1]{\csname bibitem#1\endcsname}%
\let\auto@bib@innerbib\@empty
\bibitem [{\citenamefont {Kane}\ and\ \citenamefont {Mele}(2005)}]{Kane2005}%
  \BibitemOpen
  \bibfield  {author} {\bibinfo {author} {\bibfnamefont {C.~L.}\ \bibnamefont
  {Kane}}\ and\ \bibinfo {author} {\bibfnamefont {E.~J.}\ \bibnamefont
  {Mele}},\ }\bibfield  {title} {\enquote {\bibinfo {title} {${Z}_{2}$
  topological order and the quantum spin hall effect},}\ }\href {\doibase
  10.1103/PhysRevLett.95.146802} {\bibfield  {journal} {\bibinfo  {journal}
  {Phys. Rev. Lett.}\ }\textbf {\bibinfo {volume} {95}},\ \bibinfo {pages}
  {146802} (\bibinfo {year} {2005})}\BibitemShut {NoStop}%
\bibitem [{\citenamefont {Fu}\ \emph {et~al.}(2007)\citenamefont {Fu},
  \citenamefont {Kane},\ and\ \citenamefont {Mele}}]{Fu2007}%
  \BibitemOpen
  \bibfield  {author} {\bibinfo {author} {\bibfnamefont {L.}~\bibnamefont
  {Fu}}, \bibinfo {author} {\bibfnamefont {C.~L.}\ \bibnamefont {Kane}}, \ and\
  \bibinfo {author} {\bibfnamefont {E.~J.}\ \bibnamefont {Mele}},\ }\bibfield
  {title} {\enquote {\bibinfo {title} {Topological insulators in three
  dimensions},}\ }\href {\doibase 10.1103/PhysRevLett.98.106803} {\bibfield
  {journal} {\bibinfo  {journal} {Phys. Rev. Lett.}\ }\textbf {\bibinfo
  {volume} {98}},\ \bibinfo {pages} {106803} (\bibinfo {year}
  {2007})}\BibitemShut {NoStop}%
\bibitem [{\citenamefont {Fu}\ and\ \citenamefont {Kane}(2007)}]{Fu2007-2}%
  \BibitemOpen
  \bibfield  {author} {\bibinfo {author} {\bibfnamefont {L.}~\bibnamefont
  {Fu}}\ and\ \bibinfo {author} {\bibfnamefont {C.~L.}\ \bibnamefont {Kane}},\
  }\bibfield  {title} {\enquote {\bibinfo {title} {Topological insulators with
  inversion symmetry},}\ }\href {\doibase 10.1103/PhysRevB.76.045302}
  {\bibfield  {journal} {\bibinfo  {journal} {Phys. Rev. B}\ }\textbf {\bibinfo
  {volume} {76}},\ \bibinfo {pages} {045302} (\bibinfo {year}
  {2007})}\BibitemShut {NoStop}%
\bibitem [{\citenamefont {Hasan}\ and\ \citenamefont {Kane}(2010)}]{Hasan2010}%
  \BibitemOpen
  \bibfield  {author} {\bibinfo {author} {\bibfnamefont {M.~Z.}\ \bibnamefont
  {Hasan}}\ and\ \bibinfo {author} {\bibfnamefont {C.~L.}\ \bibnamefont
  {Kane}},\ }\bibfield  {title} {\enquote {\bibinfo {title} {Colloquium:
  Topological insulators},}\ }\href {\doibase 10.1103/RevModPhys.82.3045}
  {\bibfield  {journal} {\bibinfo  {journal} {Rev. Mod. Phys.}\ }\textbf
  {\bibinfo {volume} {82}},\ \bibinfo {pages} {3045--3067} (\bibinfo {year}
  {2010})}\BibitemShut {NoStop}%
\bibitem [{\citenamefont {Shen}(2013)}]{ShenBook}%
  \BibitemOpen
  \bibfield  {author} {\bibinfo {author} {\bibfnamefont {S.-Q.}\ \bibnamefont
  {Shen}},\ }\href@noop {} {\emph {\bibinfo {title} {Topological Insulators:
  Dirac Equation in Condensed Matters}}}\ (\bibinfo  {publisher} {Springer
  Science \& Business, New York},\ \bibinfo {year} {2013})\BibitemShut
  {NoStop}%
\bibitem [{\citenamefont {Bernevig}\ and\ \citenamefont
  {Hughes}(2013)}]{BernevigBook}%
  \BibitemOpen
  \bibfield  {author} {\bibinfo {author} {\bibfnamefont {A.}~\bibnamefont
  {Bernevig}}\ and\ \bibinfo {author} {\bibfnamefont {T.~L.}\ \bibnamefont
  {Hughes}},\ }\href@noop {} {\emph {\bibinfo {title} {Topological Insulators
  and Topological Superconductors}}}\ (\bibinfo  {publisher} {Princeton
  University Press, Princeton, NJ},\ \bibinfo {year} {2013})\BibitemShut
  {NoStop}%
\bibitem [{\citenamefont {Schnyder}\ \emph {et~al.}(2008)\citenamefont
  {Schnyder}, \citenamefont {Ryu}, \citenamefont {Furusaki},\ and\
  \citenamefont {Ludwig}}]{Schnyder2008}%
  \BibitemOpen
  \bibfield  {author} {\bibinfo {author} {\bibfnamefont {A.~P.}\ \bibnamefont
  {Schnyder}}, \bibinfo {author} {\bibfnamefont {S.}~\bibnamefont {Ryu}},
  \bibinfo {author} {\bibfnamefont {A.}~\bibnamefont {Furusaki}}, \ and\
  \bibinfo {author} {\bibfnamefont {A.~W.~W.}\ \bibnamefont {Ludwig}},\
  }\bibfield  {title} {\enquote {\bibinfo {title} {Classification of
  topological insulators and superconductors in three spatial dimensions},}\
  }\href {http://link.aps.org/doi/10.1103/PhysRevB.78.195125} {\bibfield
  {journal} {\bibinfo  {journal} {Phys. Rev. B}\ }\textbf {\bibinfo {volume}
  {78}},\ \bibinfo {pages} {195125} (\bibinfo {year} {2008})}\BibitemShut
  {NoStop}%
\bibitem [{\citenamefont {Kitaev}(2009)}]{Kitaev2009}%
  \BibitemOpen
  \bibfield  {author} {\bibinfo {author} {\bibfnamefont {A.}~\bibnamefont
  {Kitaev}},\ }\bibfield  {title} {\enquote {\bibinfo {title} {Periodic table
  for topological insulators and superconductors},}\ }\href
  {https://aip.scitation.org/doi/abs/10.1063/1.3149495} {\bibfield  {journal}
  {\bibinfo  {journal} {AIP Conference Proceedings}\ }\textbf {\bibinfo
  {volume} {1134}},\ \bibinfo {pages} {22--30} (\bibinfo {year}
  {2009})}\BibitemShut {NoStop}%
\bibitem [{\citenamefont {Chiu}\ \emph {et~al.}(2016)\citenamefont {Chiu},
  \citenamefont {Teo}, \citenamefont {Schnyder},\ and\ \citenamefont
  {Ryu}}]{Chiu2016}%
  \BibitemOpen
  \bibfield  {author} {\bibinfo {author} {\bibfnamefont {C.-K.}\ \bibnamefont
  {Chiu}}, \bibinfo {author} {\bibfnamefont {J.~C.~Y.}\ \bibnamefont {Teo}},
  \bibinfo {author} {\bibfnamefont {A.~P.}\ \bibnamefont {Schnyder}}, \ and\
  \bibinfo {author} {\bibfnamefont {S.}~\bibnamefont {Ryu}},\ }\bibfield
  {title} {\enquote {\bibinfo {title} {Classification of topological quantum
  matter with symmetries},}\ }\href
  {http://link.aps.org/doi/10.1103/RevModPhys.88.035005} {\bibfield  {journal}
  {\bibinfo  {journal} {Rev. Mod. Phys.}\ }\textbf {\bibinfo {volume} {88}},\
  \bibinfo {pages} {035005} (\bibinfo {year} {2016})}\BibitemShut {NoStop}%
\bibitem [{\citenamefont {Kruthoff}\ \emph {et~al.}(2017)\citenamefont
  {Kruthoff}, \citenamefont {de~Boer}, \citenamefont {van Wezel}, \citenamefont
  {Kane},\ and\ \citenamefont {Slager}}]{Kruthoff2017}%
  \BibitemOpen
  \bibfield  {author} {\bibinfo {author} {\bibfnamefont {J.}~\bibnamefont
  {Kruthoff}}, \bibinfo {author} {\bibfnamefont {J.}~\bibnamefont {de~Boer}},
  \bibinfo {author} {\bibfnamefont {J.}~\bibnamefont {van Wezel}}, \bibinfo
  {author} {\bibfnamefont {C.~L.}\ \bibnamefont {Kane}}, \ and\ \bibinfo
  {author} {\bibfnamefont {R.-J.}\ \bibnamefont {Slager}},\ }\bibfield  {title}
  {\enquote {\bibinfo {title} {Topological classification of crystalline
  insulators through band structure combinatorics},}\ }\href {\doibase
  10.1103/PhysRevX.7.041069} {\bibfield  {journal} {\bibinfo  {journal} {Phys.
  Rev. X}\ }\textbf {\bibinfo {volume} {7}},\ \bibinfo {pages} {041069}
  (\bibinfo {year} {2017})}\BibitemShut {NoStop}%
\bibitem [{\citenamefont {Bradlyn}\ \emph {et~al.}(2017)\citenamefont
  {Bradlyn}, \citenamefont {Elcoro}, \citenamefont {Cano}, \citenamefont
  {Vergniory}, \citenamefont {Wang}, \citenamefont {Felser}, \citenamefont
  {Aroyo},\ and\ \citenamefont {Bernevig}}]{Bradlyn2017}%
  \BibitemOpen
  \bibfield  {author} {\bibinfo {author} {\bibfnamefont {B.}~\bibnamefont
  {Bradlyn}}, \bibinfo {author} {\bibfnamefont {L.}~\bibnamefont {Elcoro}},
  \bibinfo {author} {\bibfnamefont {J.}~\bibnamefont {Cano}}, \bibinfo {author}
  {\bibfnamefont {M.~G.}\ \bibnamefont {Vergniory}}, \bibinfo {author}
  {\bibfnamefont {Z.}~\bibnamefont {Wang}}, \bibinfo {author} {\bibfnamefont
  {C.}~\bibnamefont {Felser}}, \bibinfo {author} {\bibfnamefont {M.~I.}\
  \bibnamefont {Aroyo}}, \ and\ \bibinfo {author} {\bibfnamefont {B.~A.}\
  \bibnamefont {Bernevig}},\ }\bibfield  {title} {\enquote {\bibinfo {title}
  {Topological quantum chemistry},}\ }\href
  {https://doi.org/10.1038/nature23268} {\bibfield  {journal} {\bibinfo
  {journal} {Nature}\ }\textbf {\bibinfo {volume} {547}},\ \bibinfo {pages}
  {298--305} (\bibinfo {year} {2017})}\BibitemShut {NoStop}%
\bibitem [{\citenamefont {Cano}\ \emph {et~al.}(2018)\citenamefont {Cano},
  \citenamefont {Bradlyn}, \citenamefont {Wang}, \citenamefont {Elcoro},
  \citenamefont {Vergniory}, \citenamefont {Felser}, \citenamefont {Aroyo},\
  and\ \citenamefont {Bernevig}}]{Cano2018}%
  \BibitemOpen
  \bibfield  {author} {\bibinfo {author} {\bibfnamefont {J.}~\bibnamefont
  {Cano}}, \bibinfo {author} {\bibfnamefont {B.}~\bibnamefont {Bradlyn}},
  \bibinfo {author} {\bibfnamefont {Z.}~\bibnamefont {Wang}}, \bibinfo {author}
  {\bibfnamefont {L.}~\bibnamefont {Elcoro}}, \bibinfo {author} {\bibfnamefont
  {M.~G.}\ \bibnamefont {Vergniory}}, \bibinfo {author} {\bibfnamefont
  {C.}~\bibnamefont {Felser}}, \bibinfo {author} {\bibfnamefont {M.~I.}\
  \bibnamefont {Aroyo}}, \ and\ \bibinfo {author} {\bibfnamefont {B.~A.}\
  \bibnamefont {Bernevig}},\ }\bibfield  {title} {\enquote {\bibinfo {title}
  {Building blocks of topological quantum chemistry: Elementary band
  representations},}\ }\href {\doibase 10.1103/PhysRevB.97.035139} {\bibfield
  {journal} {\bibinfo  {journal} {Phys. Rev. B}\ }\textbf {\bibinfo {volume}
  {97}},\ \bibinfo {pages} {035139} (\bibinfo {year} {2018})}\BibitemShut
  {NoStop}%
\bibitem [{\citenamefont {Asb\'{o}th}\ \emph {et~al.}(2016)\citenamefont
  {Asb\'{o}th}, \citenamefont {Oroszl\'{a}ni},\ and\ \citenamefont
  {P\'{a}lyi}}]{AsbothBook}%
  \BibitemOpen
  \bibfield  {author} {\bibinfo {author} {\bibfnamefont {J.~K.}\ \bibnamefont
  {Asb\'{o}th}}, \bibinfo {author} {\bibfnamefont {L.}~\bibnamefont
  {Oroszl\'{a}ni}}, \ and\ \bibinfo {author} {\bibnamefont {P\'{a}lyi}},\
  }\href@noop {} {\emph {\bibinfo {title} {A Short Course on Topological
  Insulators}}}\ (\bibinfo  {publisher} {Springer, New York},\ \bibinfo {year}
  {2016})\BibitemShut {NoStop}%
\bibitem [{\citenamefont {Ortmann}\ \emph {et~al.}(2015)\citenamefont
  {Ortmann}, \citenamefont {Roche}, \citenamefont {Valenzuela},\ and\
  \citenamefont {Molenkamp}}]{OrtmannBook}%
  \BibitemOpen
  \bibfield  {author} {\bibinfo {author} {\bibfnamefont {F.}~\bibnamefont
  {Ortmann}}, \bibinfo {author} {\bibfnamefont {S.}~\bibnamefont {Roche}},
  \bibinfo {author} {\bibfnamefont {S.~O.}\ \bibnamefont {Valenzuela}}, \ and\
  \bibinfo {author} {\bibfnamefont {L.~W.}\ \bibnamefont {Molenkamp}},\
  }\href@noop {} {\emph {\bibinfo {title} {Topological Insulators: Fundamentals
  and Perspectives}}}\ (\bibinfo  {publisher} {Wiley, New York},\ \bibinfo
  {year} {2015})\BibitemShut {NoStop}%
\bibitem [{\citenamefont {Benalcazar}\ \emph {et~al.}(2017)\citenamefont
  {Benalcazar}, \citenamefont {Bernevig},\ and\ \citenamefont
  {Hughes}}]{Benalcazar2017}%
  \BibitemOpen
  \bibfield  {author} {\bibinfo {author} {\bibfnamefont {W.~A.}\ \bibnamefont
  {Benalcazar}}, \bibinfo {author} {\bibfnamefont {B.~A.}\ \bibnamefont
  {Bernevig}}, \ and\ \bibinfo {author} {\bibfnamefont {T.~L.}\ \bibnamefont
  {Hughes}},\ }\bibfield  {title} {\enquote {\bibinfo {title} {Quantized
  electric multipole insulators},}\ }\href {\doibase 10.1126/science.aah6442}
  {\bibfield  {journal} {\bibinfo  {journal} {Science}\ }\textbf {\bibinfo
  {volume} {357}},\ \bibinfo {pages} {61--66} (\bibinfo {year}
  {2017})}\BibitemShut {NoStop}%
\bibitem [{\citenamefont {Schindler}\ \emph {et~al.}(2018)\citenamefont
  {Schindler}, \citenamefont {Cook}, \citenamefont {Vergniory}, \citenamefont
  {Wang}, \citenamefont {Parkin}, \citenamefont {Bernevig},\ and\ \citenamefont
  {Neupert}}]{Schindler2018}%
  \BibitemOpen
  \bibfield  {author} {\bibinfo {author} {\bibfnamefont {F.}~\bibnamefont
  {Schindler}}, \bibinfo {author} {\bibfnamefont {A.~M.}\ \bibnamefont {Cook}},
  \bibinfo {author} {\bibfnamefont {M.~G.}\ \bibnamefont {Vergniory}}, \bibinfo
  {author} {\bibfnamefont {Z.}~\bibnamefont {Wang}}, \bibinfo {author}
  {\bibfnamefont {S.~S.~P.}\ \bibnamefont {Parkin}}, \bibinfo {author}
  {\bibfnamefont {B.~A.}\ \bibnamefont {Bernevig}}, \ and\ \bibinfo {author}
  {\bibfnamefont {T.}~\bibnamefont {Neupert}},\ }\bibfield  {title} {\enquote
  {\bibinfo {title} {Higher-order topological insulators},}\ }\href
  {http://advances.sciencemag.org/content/4/6/eaat0346} {\bibfield  {journal}
  {\bibinfo  {journal} {Sci. Adv.}\ }\textbf {\bibinfo {volume} {4}} (\bibinfo
  {year} {2018})}\BibitemShut {NoStop}%
\bibitem [{\citenamefont {Gong}\ \emph {et~al.}(2018)\citenamefont {Gong},
  \citenamefont {Ashida}, \citenamefont {Kawabata}, \citenamefont {Takasan},
  \citenamefont {Higashikawa},\ and\ \citenamefont {Ueda}}]{Gong2018}%
  \BibitemOpen
  \bibfield  {author} {\bibinfo {author} {\bibfnamefont {Z.}~\bibnamefont
  {Gong}}, \bibinfo {author} {\bibfnamefont {Y.}~\bibnamefont {Ashida}},
  \bibinfo {author} {\bibfnamefont {K.}~\bibnamefont {Kawabata}}, \bibinfo
  {author} {\bibfnamefont {K.}~\bibnamefont {Takasan}}, \bibinfo {author}
  {\bibfnamefont {S.}~\bibnamefont {Higashikawa}}, \ and\ \bibinfo {author}
  {\bibfnamefont {M.}~\bibnamefont {Ueda}},\ }\bibfield  {title} {\enquote
  {\bibinfo {title} {Topological phases of non-hermitian systems},}\ }\href
  {\doibase 10.1103/PhysRevX.8.031079} {\bibfield  {journal} {\bibinfo
  {journal} {Phys. Rev. X}\ }\textbf {\bibinfo {volume} {8}},\ \bibinfo {pages}
  {031079} (\bibinfo {year} {2018})}\BibitemShut {NoStop}%
\bibitem [{\citenamefont {Liu}\ \emph {et~al.}(2019)\citenamefont {Liu},
  \citenamefont {Jiang},\ and\ \citenamefont {Chen}}]{Liu2018}%
  \BibitemOpen
  \bibfield  {author} {\bibinfo {author} {\bibfnamefont {C.-H.}\ \bibnamefont
  {Liu}}, \bibinfo {author} {\bibfnamefont {H.}~\bibnamefont {Jiang}}, \ and\
  \bibinfo {author} {\bibfnamefont {S.}~\bibnamefont {Chen}},\ }\bibfield
  {title} {\enquote {\bibinfo {title} {Topological classification of
  non-hermitian systems with reflection symmetry},}\ }\href {\doibase
  10.1103/PhysRevB.99.125103} {\bibfield  {journal} {\bibinfo  {journal} {Phys.
  Rev. B}\ }\textbf {\bibinfo {volume} {99}},\ \bibinfo {pages} {125103}
  (\bibinfo {year} {2019})}\BibitemShut {NoStop}%
\bibitem [{\citenamefont {Chuang}\ and\ \citenamefont
  {Nielsen}(2000)}]{ChuangBook}%
  \BibitemOpen
  \bibfield  {author} {\bibinfo {author} {\bibfnamefont {I.}~\bibnamefont
  {Chuang}}\ and\ \bibinfo {author} {\bibfnamefont {M.}~\bibnamefont
  {Nielsen}},\ }\href@noop {} {\emph {\bibinfo {title} {Quantum Computation and
  Quantum Information}}}\ (\bibinfo  {publisher} {Cambridge University Press,
  Cambridge},\ \bibinfo {year} {2000})\BibitemShut {NoStop}%
\bibitem [{\citenamefont {Lu}\ \emph {et~al.}(2014)\citenamefont {Lu},
  \citenamefont {Joannopoulos},\ and\ \citenamefont {Soljačić}}]{Lu2014}%
  \BibitemOpen
  \bibfield  {author} {\bibinfo {author} {\bibfnamefont {L.}~\bibnamefont
  {Lu}}, \bibinfo {author} {\bibfnamefont {J.~D.}\ \bibnamefont
  {Joannopoulos}}, \ and\ \bibinfo {author} {\bibfnamefont {M.}~\bibnamefont
  {Soljačić}},\ }\bibfield  {title} {\enquote {\bibinfo {title} {Topological
  photonics},}\ }\href {https://doi.org/10.1038/nphoton.2014.248} {\bibfield
  {journal} {\bibinfo  {journal} {Nature Photonics}\ }\textbf {\bibinfo
  {volume} {8}},\ \bibinfo {pages} {821--} (\bibinfo {year}
  {2014})}\BibitemShut {NoStop}%
\bibitem [{\citenamefont {Parto}\ \emph {et~al.}(2018)\citenamefont {Parto},
  \citenamefont {Wittek}, \citenamefont {Hodaei}, \citenamefont {Harari},
  \citenamefont {Bandres}, \citenamefont {Ren}, \citenamefont {Rechtsman},
  \citenamefont {Segev}, \citenamefont {Christodoulides},\ and\ \citenamefont
  {Khajavikhan}}]{Parto2018}%
  \BibitemOpen
  \bibfield  {author} {\bibinfo {author} {\bibfnamefont {M.}~\bibnamefont
  {Parto}}, \bibinfo {author} {\bibfnamefont {S.}~\bibnamefont {Wittek}},
  \bibinfo {author} {\bibfnamefont {H.}~\bibnamefont {Hodaei}}, \bibinfo
  {author} {\bibfnamefont {G.}~\bibnamefont {Harari}}, \bibinfo {author}
  {\bibfnamefont {M.~A.}\ \bibnamefont {Bandres}}, \bibinfo {author}
  {\bibfnamefont {J.}~\bibnamefont {Ren}}, \bibinfo {author} {\bibfnamefont
  {M.~C.}\ \bibnamefont {Rechtsman}}, \bibinfo {author} {\bibfnamefont
  {M.}~\bibnamefont {Segev}}, \bibinfo {author} {\bibfnamefont {D.~N.}\
  \bibnamefont {Christodoulides}}, \ and\ \bibinfo {author} {\bibfnamefont
  {M.}~\bibnamefont {Khajavikhan}},\ }\bibfield  {title} {\enquote {\bibinfo
  {title} {Edge-mode lasing in 1d topological active arrays},}\ }\href
  {\doibase 10.1103/PhysRevLett.120.113901} {\bibfield  {journal} {\bibinfo
  {journal} {Phys. Rev. Lett.}\ }\textbf {\bibinfo {volume} {120}},\ \bibinfo
  {pages} {113901} (\bibinfo {year} {2018})}\BibitemShut {NoStop}%
\bibitem [{\citenamefont {Takata}\ and\ \citenamefont
  {Notomi}(2018)}]{Takata2018}%
  \BibitemOpen
  \bibfield  {author} {\bibinfo {author} {\bibfnamefont {K.}~\bibnamefont
  {Takata}}\ and\ \bibinfo {author} {\bibfnamefont {M.}~\bibnamefont
  {Notomi}},\ }\bibfield  {title} {\enquote {\bibinfo {title} {Photonic
  topological insulating phase induced solely by gain and loss},}\ }\href
  {\doibase 10.1103/PhysRevLett.121.213902} {\bibfield  {journal} {\bibinfo
  {journal} {Phys. Rev. Lett.}\ }\textbf {\bibinfo {volume} {121}},\ \bibinfo
  {pages} {213902} (\bibinfo {year} {2018})}\BibitemShut {NoStop}%
\bibitem [{\citenamefont {{Zhu}}\ \emph {et~al.}()\citenamefont {{Zhu}},
  \citenamefont {{Gupta}}, \citenamefont {{Sun}}, \citenamefont {{He}},
  \citenamefont {{Li}}, \citenamefont {{Jiang}}, \citenamefont {{Lu}},
  \citenamefont {{Liu}},\ and\ \citenamefont {{Chen}}}]{Zhu2018}%
  \BibitemOpen
  \bibfield  {author} {\bibinfo {author} {\bibfnamefont {X.-Y.}\ \bibnamefont
  {{Zhu}}}, \bibinfo {author} {\bibfnamefont {S.~K.}\ \bibnamefont {{Gupta}}},
  \bibinfo {author} {\bibfnamefont {X.-C.}\ \bibnamefont {{Sun}}}, \bibinfo
  {author} {\bibfnamefont {C.}~\bibnamefont {{He}}}, \bibinfo {author}
  {\bibfnamefont {G.-X.}\ \bibnamefont {{Li}}}, \bibinfo {author}
  {\bibfnamefont {J.-H.}\ \bibnamefont {{Jiang}}}, \bibinfo {author}
  {\bibfnamefont {M.-H.}\ \bibnamefont {{Lu}}}, \bibinfo {author}
  {\bibfnamefont {X.-P.}\ \bibnamefont {{Liu}}}, \ and\ \bibinfo {author}
  {\bibfnamefont {Y.-F.}\ \bibnamefont {{Chen}}},\ }\bibfield  {title}
  {\enquote {\bibinfo {title} {{Topological Flat Band and Parity-Time Symmetry
  in a Honeycomb Lattice of Coupled Resonant Optical Waveguides}},}\ }\href
  {https://arxiv.org/abs/1801.10289} {\ }\Eprint
  {http://arxiv.org/abs/1801.10289} {arXiv:1801.10289} \BibitemShut {NoStop}%
\bibitem [{\citenamefont {Longhi}(2018)}]{Longhi2018}%
  \BibitemOpen
  \bibfield  {author} {\bibinfo {author} {\bibfnamefont {S.}~\bibnamefont
  {Longhi}},\ }\bibfield  {title} {\enquote {\bibinfo {title} {Non-hermitian
  gauged topological laser arrays},}\ }\href {\doibase 10.1002/andp.201800023}
  {\bibfield  {journal} {\bibinfo  {journal} {Annalen der Physik}\ }\textbf
  {\bibinfo {volume} {530}},\ \bibinfo {pages} {1800023} (\bibinfo {year}
  {2018})}\BibitemShut {NoStop}%
\bibitem [{\citenamefont {{Kozii}}\ and\ \citenamefont {{Fu}}()}]{Kozii2017}%
  \BibitemOpen
  \bibfield  {author} {\bibinfo {author} {\bibfnamefont {V.}~\bibnamefont
  {{Kozii}}}\ and\ \bibinfo {author} {\bibfnamefont {L.}~\bibnamefont {{Fu}}},\
  }\bibfield  {title} {\enquote {\bibinfo {title} {{Non-Hermitian Topological
  Theory of Finite-Lifetime Quasiparticles: Prediction of Bulk Fermi Arc Due to
  Exceptional Point}},}\ }\href {https://arxiv.org/abs/1708.0584} {\ }\Eprint
  {http://arxiv.org/abs/1708.05841} {arXiv:1708.05841} \BibitemShut {NoStop}%
\bibitem [{\citenamefont {Yoshida}\ \emph {et~al.}(2018)\citenamefont
  {Yoshida}, \citenamefont {Peters},\ and\ \citenamefont
  {Kawakami}}]{Yoshida2018}%
  \BibitemOpen
  \bibfield  {author} {\bibinfo {author} {\bibfnamefont {T.}~\bibnamefont
  {Yoshida}}, \bibinfo {author} {\bibfnamefont {R.}~\bibnamefont {Peters}}, \
  and\ \bibinfo {author} {\bibfnamefont {N.}~\bibnamefont {Kawakami}},\
  }\bibfield  {title} {\enquote {\bibinfo {title} {Non-hermitian perspective of
  the band structure in heavy-fermion systems},}\ }\href {\doibase
  10.1103/PhysRevB.98.035141} {\bibfield  {journal} {\bibinfo  {journal} {Phys.
  Rev. B}\ }\textbf {\bibinfo {volume} {98}},\ \bibinfo {pages} {035141}
  (\bibinfo {year} {2018})}\BibitemShut {NoStop}%
\bibitem [{\citenamefont {Martinez~Alvarez}\ \emph {et~al.}(2018)\citenamefont
  {Martinez~Alvarez}, \citenamefont {Barrios~Vargas}, \citenamefont
  {Berdakin},\ and\ \citenamefont {Foa~Torres}}]{MartinezAlvarez2018}%
  \BibitemOpen
  \bibfield  {author} {\bibinfo {author} {\bibfnamefont {V.~M.}\ \bibnamefont
  {Martinez~Alvarez}}, \bibinfo {author} {\bibfnamefont {J.~E.}\ \bibnamefont
  {Barrios~Vargas}}, \bibinfo {author} {\bibfnamefont {M.}~\bibnamefont
  {Berdakin}}, \ and\ \bibinfo {author} {\bibfnamefont {L.~E.~F.}\ \bibnamefont
  {Foa~Torres}},\ }\bibfield  {title} {\enquote {\bibinfo {title} {Topological
  states of non-hermitian systems},}\ }\href
  {https://doi.org/10.1140/epjst/e2018-800091-5} {\bibfield  {journal}
  {\bibinfo  {journal} {Eur. Phys. J. Spec. Top.}\ }\textbf {\bibinfo {volume}
  {227}},\ \bibinfo {pages} {1295} (\bibinfo {year} {2018})}\BibitemShut
  {NoStop}%
\bibitem [{\citenamefont {Lee}(2016)}]{Lee2016}%
  \BibitemOpen
  \bibfield  {author} {\bibinfo {author} {\bibfnamefont {T.}~\bibnamefont
  {Lee}},\ }\bibfield  {title} {\enquote {\bibinfo {title} {Anomalous edge
  state in a non-hermitian lattice},}\ }\href {\doibase
  10.1103/PhysRevLett.116.133903} {\bibfield  {journal} {\bibinfo  {journal}
  {Phys. Rev. Lett.}\ }\textbf {\bibinfo {volume} {116}},\ \bibinfo {pages}
  {133903} (\bibinfo {year} {2016})}\BibitemShut {NoStop}%
\bibitem [{\citenamefont {Leykam}\ \emph {et~al.}(2017)\citenamefont {Leykam},
  \citenamefont {Bliokh}, \citenamefont {Huang}, \citenamefont {Chong},\ and\
  \citenamefont {Nori}}]{Leykam2017}%
  \BibitemOpen
  \bibfield  {author} {\bibinfo {author} {\bibfnamefont {D.}~\bibnamefont
  {Leykam}}, \bibinfo {author} {\bibfnamefont {K.~Y.}\ \bibnamefont {Bliokh}},
  \bibinfo {author} {\bibfnamefont {C.}~\bibnamefont {Huang}}, \bibinfo
  {author} {\bibfnamefont {Y.~D.}\ \bibnamefont {Chong}}, \ and\ \bibinfo
  {author} {\bibfnamefont {F.}~\bibnamefont {Nori}},\ }\bibfield  {title}
  {\enquote {\bibinfo {title} {Edge modes, degeneracies, and topological
  numbers in non-hermitian systems},}\ }\href {\doibase
  10.1103/PhysRevLett.118.040401} {\bibfield  {journal} {\bibinfo  {journal}
  {Phys. Rev. Lett.}\ }\textbf {\bibinfo {volume} {118}},\ \bibinfo {pages}
  {040401} (\bibinfo {year} {2017})}\BibitemShut {NoStop}%
\bibitem [{\citenamefont {Xiong}(2018)}]{Xiong2018}%
  \BibitemOpen
  \bibfield  {author} {\bibinfo {author} {\bibfnamefont {Y.}~\bibnamefont
  {Xiong}},\ }\bibfield  {title} {\enquote {\bibinfo {title} {Why does bulk
  boundary correspondence fail in some non-hermitian topological models},}\
  }\href {http://stacks.iop.org/2399-6528/2/i=3/a=035043} {\bibfield  {journal}
  {\bibinfo  {journal} {Journal of Physics Communications}\ }\textbf {\bibinfo
  {volume} {2}},\ \bibinfo {pages} {035043} (\bibinfo {year}
  {2018})}\BibitemShut {NoStop}%
\bibitem [{\citenamefont {Yao}\ \emph {et~al.}(2018)\citenamefont {Yao},
  \citenamefont {Song},\ and\ \citenamefont {Wang}}]{Yao2018}%
  \BibitemOpen
  \bibfield  {author} {\bibinfo {author} {\bibfnamefont {S.}~\bibnamefont
  {Yao}}, \bibinfo {author} {\bibfnamefont {F.}~\bibnamefont {Song}}, \ and\
  \bibinfo {author} {\bibfnamefont {Z.}~\bibnamefont {Wang}},\ }\bibfield
  {title} {\enquote {\bibinfo {title} {Non-hermitian chern bands},}\ }\href
  {\doibase 10.1103/PhysRevLett.121.136802} {\bibfield  {journal} {\bibinfo
  {journal} {Phys. Rev. Lett.}\ }\textbf {\bibinfo {volume} {121}},\ \bibinfo
  {pages} {136802} (\bibinfo {year} {2018})}\BibitemShut {NoStop}%
\bibitem [{\citenamefont {Yao}\ and\ \citenamefont {Wang}(2018)}]{Yao2018-2}%
  \BibitemOpen
  \bibfield  {author} {\bibinfo {author} {\bibfnamefont {S.}~\bibnamefont
  {Yao}}\ and\ \bibinfo {author} {\bibfnamefont {Z.}~\bibnamefont {Wang}},\
  }\bibfield  {title} {\enquote {\bibinfo {title} {Edge states and topological
  invariants of non-hermitian systems},}\ }\href {\doibase
  10.1103/PhysRevLett.121.086803} {\bibfield  {journal} {\bibinfo  {journal}
  {Phys. Rev. Lett.}\ }\textbf {\bibinfo {volume} {121}},\ \bibinfo {pages}
  {086803} (\bibinfo {year} {2018})}\BibitemShut {NoStop}%
\bibitem [{\citenamefont {Kunst}\ \emph {et~al.}(2018)\citenamefont {Kunst},
  \citenamefont {Edvardsson}, \citenamefont {Budich},\ and\ \citenamefont
  {Bergholtz}}]{Kunst2018}%
  \BibitemOpen
  \bibfield  {author} {\bibinfo {author} {\bibfnamefont {F.~K.}\ \bibnamefont
  {Kunst}}, \bibinfo {author} {\bibfnamefont {E.}~\bibnamefont {Edvardsson}},
  \bibinfo {author} {\bibfnamefont {J.~C.}\ \bibnamefont {Budich}}, \ and\
  \bibinfo {author} {\bibfnamefont {E.~J.}\ \bibnamefont {Bergholtz}},\
  }\bibfield  {title} {\enquote {\bibinfo {title} {Biorthogonal bulk-boundary
  correspondence in non-hermitian systems},}\ }\href {\doibase
  10.1103/PhysRevLett.121.026808} {\bibfield  {journal} {\bibinfo  {journal}
  {Phys. Rev. Lett.}\ }\textbf {\bibinfo {volume} {121}},\ \bibinfo {pages}
  {026808} (\bibinfo {year} {2018})}\BibitemShut {NoStop}%
\bibitem [{\citenamefont {Lee}\ and\ \citenamefont
  {Thomale}(2019)}]{Lee2018-3}%
  \BibitemOpen
  \bibfield  {author} {\bibinfo {author} {\bibfnamefont {C.~H.}\ \bibnamefont
  {Lee}}\ and\ \bibinfo {author} {\bibfnamefont {R.}~\bibnamefont {Thomale}},\
  }\bibfield  {title} {\enquote {\bibinfo {title} {Anatomy of skin modes and
  topology in non-hermitian systems},}\ }\href
  {https://link.aps.org/doi/10.1103/PhysRevB.99.201103} {\bibfield  {journal}
  {\bibinfo  {journal} {Phys. Rev. B}\ }\textbf {\bibinfo {volume} {99}},\
  \bibinfo {pages} {201103} (\bibinfo {year} {2019})}\BibitemShut {NoStop}%
\bibitem [{\citenamefont {Jin}\ and\ \citenamefont {Song}(2019)}]{Jin2018}%
  \BibitemOpen
  \bibfield  {author} {\bibinfo {author} {\bibfnamefont {L.}~\bibnamefont
  {Jin}}\ and\ \bibinfo {author} {\bibfnamefont {Z.}~\bibnamefont {Song}},\
  }\bibfield  {title} {\enquote {\bibinfo {title} {Bulk-boundary correspondence
  in a non-hermitian system in one dimension with chiral inversion symmetry},}\
  }\href {\doibase 10.1103/PhysRevB.99.081103} {\bibfield  {journal} {\bibinfo
  {journal} {Phys. Rev. B}\ }\textbf {\bibinfo {volume} {99}},\ \bibinfo
  {pages} {081103} (\bibinfo {year} {2019})}\BibitemShut {NoStop}%
\bibitem [{\citenamefont {{Edvardsson}}\ \emph {et~al.}()\citenamefont
  {{Edvardsson}}, \citenamefont {{Kunst}},\ and\ \citenamefont
  {{Bergholtz}}}]{Edvardsson2018}%
  \BibitemOpen
  \bibfield  {author} {\bibinfo {author} {\bibfnamefont {E.}~\bibnamefont
  {{Edvardsson}}}, \bibinfo {author} {\bibfnamefont {F.~K.}\ \bibnamefont
  {{Kunst}}}, \ and\ \bibinfo {author} {\bibfnamefont {E.~J.}\ \bibnamefont
  {{Bergholtz}}},\ }\bibfield  {title} {\enquote {\bibinfo {title}
  {{Non-Hermitian extensions of higher-order topological phases and their
  biorthogonal bulk-boundary correspondence}},}\ }\href
  {https://arxiv.org/abs/1812.09060} {\ }\Eprint
  {http://arxiv.org/abs/1812.09060} {arXiv:1812.09060} \BibitemShut {NoStop}%
\bibitem [{\citenamefont {Su}\ \emph {et~al.}(1980)\citenamefont {Su},
  \citenamefont {Schrieffer},\ and\ \citenamefont {Heeger}}]{SSH1980}%
  \BibitemOpen
  \bibfield  {author} {\bibinfo {author} {\bibfnamefont {W.~P.}\ \bibnamefont
  {Su}}, \bibinfo {author} {\bibfnamefont {J.~R.}\ \bibnamefont {Schrieffer}},
  \ and\ \bibinfo {author} {\bibfnamefont {A.~J.}\ \bibnamefont {Heeger}},\
  }\bibfield  {title} {\enquote {\bibinfo {title} {Soliton excitations in
  polyacetylene},}\ }\href {\doibase 10.1103/PhysRevB.22.2099} {\bibfield
  {journal} {\bibinfo  {journal} {Phys. Rev. B}\ }\textbf {\bibinfo {volume}
  {22}},\ \bibinfo {pages} {2099--2111} (\bibinfo {year} {1980})}\BibitemShut
  {NoStop}%
\bibitem [{\citenamefont {Yin}\ \emph {et~al.}(2018)\citenamefont {Yin},
  \citenamefont {Jiang}, \citenamefont {Li}, \citenamefont {L\"u},\ and\
  \citenamefont {Chen}}]{Yin2018}%
  \BibitemOpen
  \bibfield  {author} {\bibinfo {author} {\bibfnamefont {C.}~\bibnamefont
  {Yin}}, \bibinfo {author} {\bibfnamefont {H.}~\bibnamefont {Jiang}}, \bibinfo
  {author} {\bibfnamefont {L.}~\bibnamefont {Li}}, \bibinfo {author}
  {\bibfnamefont {R.}~\bibnamefont {L\"u}}, \ and\ \bibinfo {author}
  {\bibfnamefont {S.}~\bibnamefont {Chen}},\ }\bibfield  {title} {\enquote
  {\bibinfo {title} {Geometrical meaning of winding number and its
  characterization of topological phases in one-dimensional chiral
  non-hermitian systems},}\ }\href {\doibase 10.1103/PhysRevA.97.052115}
  {\bibfield  {journal} {\bibinfo  {journal} {Phys. Rev. A}\ }\textbf {\bibinfo
  {volume} {97}},\ \bibinfo {pages} {052115} (\bibinfo {year}
  {2018})}\BibitemShut {NoStop}%
\bibitem [{\citenamefont {Reichel}\ and\ \citenamefont
  {Trefethen}(1992)}]{Reichel1992}%
  \BibitemOpen
  \bibfield  {author} {\bibinfo {author} {\bibfnamefont {L.}~\bibnamefont
  {Reichel}}\ and\ \bibinfo {author} {\bibfnamefont {L.~N.}\ \bibnamefont
  {Trefethen}},\ }\bibfield  {title} {\enquote {\bibinfo {title} {Eigenvalues
  and pseudo-eigenvalues of toeplitz matrices},}\ }\href {\doibase
  https://doi.org/10.1016/0024-3795(92)90374-J} {\bibfield  {journal} {\bibinfo
   {journal} {Linear Algebra and its Applications}\ }\textbf {\bibinfo {volume}
  {162-164}},\ \bibinfo {pages} {153 -- 185} (\bibinfo {year}
  {1992})}\BibitemShut {NoStop}%
\bibitem [{\citenamefont {Krause}(1994)}]{Krause1994}%
  \BibitemOpen
  \bibfield  {author} {\bibinfo {author} {\bibfnamefont {G.~M.}\ \bibnamefont
  {Krause}},\ }\bibfield  {title} {\enquote {\bibinfo {title} {Bounds for the
  variation of matrix eigenvalues and polynomial roots},}\ }\href
  {http://www.sciencedirect.com/science/article/pii/0024379594904324}
  {\bibfield  {journal} {\bibinfo  {journal} {Linear Algebra and its
  Applications}\ }\textbf {\bibinfo {volume} {208-209}},\ \bibinfo {pages} {73
  -- 82} (\bibinfo {year} {1994})}\BibitemShut {NoStop}%
\bibitem [{\citenamefont {{Bernevig}}\ and\ \citenamefont
  {{Neupert}}()}]{Bernevig2015}%
  \BibitemOpen
  \bibfield  {author} {\bibinfo {author} {\bibfnamefont {A.}~\bibnamefont
  {{Bernevig}}}\ and\ \bibinfo {author} {\bibfnamefont {T.}~\bibnamefont
  {{Neupert}}},\ }\bibfield  {title} {\enquote {\bibinfo {title} {{Topological
  Superconductors and Category Theory}},}\ }\href
  {https://arxiv.org/abs/1506.05805} {\ }\Eprint
  {http://arxiv.org/abs/1506.05805} {arXiv:1506.05805} \BibitemShut {NoStop}%
\bibitem [{\citenamefont {Jiang}\ \emph {et~al.}(2018)\citenamefont {Jiang},
  \citenamefont {Yang},\ and\ \citenamefont {Chen}}]{Jiang2018}%
  \BibitemOpen
  \bibfield  {author} {\bibinfo {author} {\bibfnamefont {H.}~\bibnamefont
  {Jiang}}, \bibinfo {author} {\bibfnamefont {C.}~\bibnamefont {Yang}}, \ and\
  \bibinfo {author} {\bibfnamefont {S.}~\bibnamefont {Chen}},\ }\bibfield
  {title} {\enquote {\bibinfo {title} {Topological invariants and phase
  diagrams for one-dimensional two-band non-hermitian systems without chiral
  symmetry},}\ }\href {\doibase 10.1103/PhysRevA.98.052116} {\bibfield
  {journal} {\bibinfo  {journal} {Phys. Rev. A}\ }\textbf {\bibinfo {volume}
  {98}},\ \bibinfo {pages} {052116} (\bibinfo {year} {2018})}\BibitemShut
  {NoStop}%
\bibitem [{\citenamefont {Li}\ and\ \citenamefont {Haldane}(2008)}]{Li2008}%
  \BibitemOpen
  \bibfield  {author} {\bibinfo {author} {\bibfnamefont {H.}~\bibnamefont
  {Li}}\ and\ \bibinfo {author} {\bibfnamefont {F.~D.~M.}\ \bibnamefont
  {Haldane}},\ }\bibfield  {title} {\enquote {\bibinfo {title} {Entanglement
  spectrum as a generalization of entanglement entropy: Identification of
  topological order in non-abelian fractional quantum hall effect states},}\
  }\href {https://link.aps.org/doi/10.1103/PhysRevLett.101.010504} {\bibfield
  {journal} {\bibinfo  {journal} {Phys. Rev. Lett.}\ }\textbf {\bibinfo
  {volume} {101}},\ \bibinfo {pages} {010504} (\bibinfo {year}
  {2008})}\BibitemShut {NoStop}%
\bibitem [{\citenamefont {Pollmann}\ \emph {et~al.}(2010)\citenamefont
  {Pollmann}, \citenamefont {Turner}, \citenamefont {Berg},\ and\ \citenamefont
  {Oshikawa}}]{Pollmann2010}%
  \BibitemOpen
  \bibfield  {author} {\bibinfo {author} {\bibfnamefont {F.}~\bibnamefont
  {Pollmann}}, \bibinfo {author} {\bibfnamefont {A.~M.}\ \bibnamefont
  {Turner}}, \bibinfo {author} {\bibfnamefont {E.}~\bibnamefont {Berg}}, \ and\
  \bibinfo {author} {\bibfnamefont {M.}~\bibnamefont {Oshikawa}},\ }\bibfield
  {title} {\enquote {\bibinfo {title} {Entanglement spectrum of a topological
  phase in one dimension},}\ }\href
  {https://link.aps.org/doi/10.1103/PhysRevB.81.064439} {\bibfield  {journal}
  {\bibinfo  {journal} {Phys. Rev. B}\ }\textbf {\bibinfo {volume} {81}},\
  \bibinfo {pages} {064439} (\bibinfo {year} {2010})}\BibitemShut {NoStop}%
\bibitem [{\citenamefont {Chandran}\ \emph {et~al.}(2011)\citenamefont
  {Chandran}, \citenamefont {Hermanns}, \citenamefont {Regnault},\ and\
  \citenamefont {Bernevig}}]{Chandran2011}%
  \BibitemOpen
  \bibfield  {author} {\bibinfo {author} {\bibfnamefont {A.}~\bibnamefont
  {Chandran}}, \bibinfo {author} {\bibfnamefont {M.}~\bibnamefont {Hermanns}},
  \bibinfo {author} {\bibfnamefont {N.}~\bibnamefont {Regnault}}, \ and\
  \bibinfo {author} {\bibfnamefont {B.~A.}\ \bibnamefont {Bernevig}},\
  }\bibfield  {title} {\enquote {\bibinfo {title} {Bulk-edge correspondence in
  entanglement spectra},}\ }\href {\doibase 10.1103/PhysRevB.84.205136}
  {\bibfield  {journal} {\bibinfo  {journal} {Phys. Rev. B}\ }\textbf {\bibinfo
  {volume} {84}},\ \bibinfo {pages} {205136} (\bibinfo {year}
  {2011})}\BibitemShut {NoStop}%
\bibitem [{\citenamefont {Qi}\ \emph {et~al.}(2012)\citenamefont {Qi},
  \citenamefont {Katsura},\ and\ \citenamefont {Ludwig}}]{Qi2012}%
  \BibitemOpen
  \bibfield  {author} {\bibinfo {author} {\bibfnamefont {X.-L.}\ \bibnamefont
  {Qi}}, \bibinfo {author} {\bibfnamefont {H.}~\bibnamefont {Katsura}}, \ and\
  \bibinfo {author} {\bibfnamefont {A.~W.~W.}\ \bibnamefont {Ludwig}},\
  }\bibfield  {title} {\enquote {\bibinfo {title} {General relationship between
  the entanglement spectrum and the edge state spectrum of topological quantum
  states},}\ }\href {https://link.aps.org/doi/10.1103/PhysRevLett.108.196402}
  {\bibfield  {journal} {\bibinfo  {journal} {Phys. Rev. Lett.}\ }\textbf
  {\bibinfo {volume} {108}},\ \bibinfo {pages} {196402} (\bibinfo {year}
  {2012})}\BibitemShut {NoStop}%
\bibitem [{\citenamefont {Peschel}(2003)}]{Peschel2003}%
  \BibitemOpen
  \bibfield  {author} {\bibinfo {author} {\bibfnamefont {I.}~\bibnamefont
  {Peschel}},\ }\bibfield  {title} {\enquote {\bibinfo {title} {Calculation of
  reduced density matrices from correlation functions},}\ }\href
  {http://stacks.iop.org/0305-4470/36/i=14/a=101} {\bibfield  {journal}
  {\bibinfo  {journal} {Journal of Physics A: Mathematical and General}\
  }\textbf {\bibinfo {volume} {36}},\ \bibinfo {pages} {L205} (\bibinfo {year}
  {2003})}\BibitemShut {NoStop}%
\bibitem [{Note1()}]{Note1}%
  \BibitemOpen
  \bibinfo {note} {A shift by the identity of a non-Hermitian Hamiltonian is
  not trivial for its flattened representation $Q$}\BibitemShut {NoStop}%
\bibitem [{\citenamefont {Shen}\ \emph {et~al.}(2018)\citenamefont {Shen},
  \citenamefont {Zhen},\ and\ \citenamefont {Fu}}]{Shen2018}%
  \BibitemOpen
  \bibfield  {author} {\bibinfo {author} {\bibfnamefont {H.}~\bibnamefont
  {Shen}}, \bibinfo {author} {\bibfnamefont {B.}~\bibnamefont {Zhen}}, \ and\
  \bibinfo {author} {\bibfnamefont {L.}~\bibnamefont {Fu}},\ }\bibfield
  {title} {\enquote {\bibinfo {title} {Topological band theory for
  non-hermitian hamiltonians},}\ }\href {\doibase
  10.1103/PhysRevLett.120.146402} {\bibfield  {journal} {\bibinfo  {journal}
  {Phys. Rev. Lett.}\ }\textbf {\bibinfo {volume} {120}},\ \bibinfo {pages}
  {146402} (\bibinfo {year} {2018})}\BibitemShut {NoStop}%
\bibitem [{\citenamefont {Turner}\ \emph {et~al.}(2010)\citenamefont {Turner},
  \citenamefont {Zhang},\ and\ \citenamefont {Vishwanath}}]{Turner2010}%
  \BibitemOpen
  \bibfield  {author} {\bibinfo {author} {\bibfnamefont {A.~M.}\ \bibnamefont
  {Turner}}, \bibinfo {author} {\bibfnamefont {Y.}~\bibnamefont {Zhang}}, \
  and\ \bibinfo {author} {\bibfnamefont {A.}~\bibnamefont {Vishwanath}},\
  }\bibfield  {title} {\enquote {\bibinfo {title} {Entanglement and inversion
  symmetry in topological insulators},}\ }\href
  {https://link.aps.org/doi/10.1103/PhysRevB.82.241102} {\bibfield  {journal}
  {\bibinfo  {journal} {Phys. Rev. B}\ }\textbf {\bibinfo {volume} {82}},\
  \bibinfo {pages} {241102} (\bibinfo {year} {2010})}\BibitemShut {NoStop}%
\bibitem [{\citenamefont {Niu}\ \emph {et~al.}(1985)\citenamefont {Niu},
  \citenamefont {Thouless},\ and\ \citenamefont {Wu}}]{Niu1985}%
  \BibitemOpen
  \bibfield  {author} {\bibinfo {author} {\bibfnamefont {Q.}~\bibnamefont
  {Niu}}, \bibinfo {author} {\bibfnamefont {D.~J.}\ \bibnamefont {Thouless}}, \
  and\ \bibinfo {author} {\bibfnamefont {Y.-S.}\ \bibnamefont {Wu}},\
  }\bibfield  {title} {\enquote {\bibinfo {title} {Quantized hall conductance
  as a topological invariant},}\ }\href {\doibase 10.1103/PhysRevB.31.3372}
  {\bibfield  {journal} {\bibinfo  {journal} {Phys. Rev. B}\ }\textbf {\bibinfo
  {volume} {31}},\ \bibinfo {pages} {3372--3377} (\bibinfo {year}
  {1985})}\BibitemShut {NoStop}%
\bibitem [{\citenamefont {Prodan}\ \emph {et~al.}(2010)\citenamefont {Prodan},
  \citenamefont {Hughes},\ and\ \citenamefont {Bernevig}}]{Prodan2010}%
  \BibitemOpen
  \bibfield  {author} {\bibinfo {author} {\bibfnamefont {E.}~\bibnamefont
  {Prodan}}, \bibinfo {author} {\bibfnamefont {T.~L.}\ \bibnamefont {Hughes}},
  \ and\ \bibinfo {author} {\bibfnamefont {B.~A.}\ \bibnamefont {Bernevig}},\
  }\bibfield  {title} {\enquote {\bibinfo {title} {Entanglement spectrum of a
  disordered topological chern insulator},}\ }\href
  {https://link.aps.org/doi/10.1103/PhysRevLett.105.115501} {\bibfield
  {journal} {\bibinfo  {journal} {Phys. Rev. Lett.}\ }\textbf {\bibinfo
  {volume} {105}},\ \bibinfo {pages} {115501} (\bibinfo {year}
  {2010})}\BibitemShut {NoStop}%
\bibitem [{\citenamefont {Mondragon-Shem}\ \emph {et~al.}(2014)\citenamefont
  {Mondragon-Shem}, \citenamefont {Hughes}, \citenamefont {Song},\ and\
  \citenamefont {Prodan}}]{Mondragon2014}%
  \BibitemOpen
  \bibfield  {author} {\bibinfo {author} {\bibfnamefont {I.}~\bibnamefont
  {Mondragon-Shem}}, \bibinfo {author} {\bibfnamefont {T.~L.}\ \bibnamefont
  {Hughes}}, \bibinfo {author} {\bibfnamefont {J.}~\bibnamefont {Song}}, \ and\
  \bibinfo {author} {\bibfnamefont {E.}~\bibnamefont {Prodan}},\ }\bibfield
  {title} {\enquote {\bibinfo {title} {Topological criticality in the
  chiral-symmetric aiii class at strong disorder},}\ }\href
  {https://link.aps.org/doi/10.1103/PhysRevLett.113.046802} {\bibfield
  {journal} {\bibinfo  {journal} {Phys. Rev. Lett.}\ }\textbf {\bibinfo
  {volume} {113}},\ \bibinfo {pages} {046802} (\bibinfo {year}
  {2014})}\BibitemShut {NoStop}%
\bibitem [{\citenamefont {Porras}\ and\ \citenamefont
  {Fern\'andez-Lorenzo}(2019)}]{Porras2018}%
  \BibitemOpen
  \bibfield  {author} {\bibinfo {author} {\bibfnamefont {D.}~\bibnamefont
  {Porras}}\ and\ \bibinfo {author} {\bibfnamefont {S.}~\bibnamefont
  {Fern\'andez-Lorenzo}},\ }\bibfield  {title} {\enquote {\bibinfo {title}
  {Topological amplification in photonic lattices},}\ }\href {\doibase
  10.1103/PhysRevLett.122.143901} {\bibfield  {journal} {\bibinfo  {journal}
  {Phys. Rev. Lett.}\ }\textbf {\bibinfo {volume} {122}},\ \bibinfo {pages}
  {143901} (\bibinfo {year} {2019})}\BibitemShut {NoStop}%
\bibitem [{\citenamefont {{Kawabata}}\ \emph {et~al.}()\citenamefont
  {{Kawabata}}, \citenamefont {{Shiozaki}}, \citenamefont {{Ueda}},\ and\
  \citenamefont {{Sato}}}]{Kawabata2018}%
  \BibitemOpen
  \bibfield  {author} {\bibinfo {author} {\bibfnamefont {K.}~\bibnamefont
  {{Kawabata}}}, \bibinfo {author} {\bibfnamefont {K.}~\bibnamefont
  {{Shiozaki}}}, \bibinfo {author} {\bibfnamefont {M.}~\bibnamefont {{Ueda}}},
  \ and\ \bibinfo {author} {\bibfnamefont {M.}~\bibnamefont {{Sato}}},\
  }\bibfield  {title} {\enquote {\bibinfo {title} {{Symmetry and Topology in
  Non-Hermitian Physics}},}\ }\href {https://arxiv.org/abs/1812.09133} {\
  }\Eprint {http://arxiv.org/abs/1812.09133} {arXiv:1812.09133} \BibitemShut
  {NoStop}%
\bibitem [{\citenamefont {{Zhou}}\ and\ \citenamefont {{Lee}}()}]{Zhou2018}%
  \BibitemOpen
  \bibfield  {author} {\bibinfo {author} {\bibfnamefont {H.}~\bibnamefont
  {{Zhou}}}\ and\ \bibinfo {author} {\bibfnamefont {J.~Y.}\ \bibnamefont
  {{Lee}}},\ }\bibfield  {title} {\enquote {\bibinfo {title} {{Periodic Table
  for Topological Bands with Non-Hermitian Bernard-LeClair Symmetries}},}\
  }\href {https://arxiv.org/abs/1812.10490} {\ }\Eprint
  {http://arxiv.org/abs/1812.10490} {arXiv:1812.10490} \BibitemShut {NoStop}%
\end{thebibliography}%

\end{document}